\documentclass[pra,letterpaper,twocolumn,showpacs,superscriptaddress,floatfix]{revtex4}
\usepackage{graphicx,psfrag,amsmath,amssymb,amsfonts,bbm,latexsym,color,dcolumn,bm,mathbbol,mathrsfs,longtable}

\def\d{\mathrm{d}}

\newcommand{\scp}[2]{\langle #1 , #2 \rangle}
\newcommand{\trace}[1]{\tr\left( #1 \right)}
\newcommand{\ba}{\begin{eqnarray}}
\newcommand{\ea}{\end{eqnarray}}
\newcommand{\bary}{\begin{array}}
\newcommand{\ear}{\end{array}}

\newcommand{\tr}{\mathop{\mathrm{tr}}\limits}
\def\IM{\mathop{\rm Im}} %

\begin{document}

\title{ Spontaneous symmetry breaking and inversion-line spectroscopy
  in gas mixtures}

\author{Carlo Presilla} \affiliation{Dipartimento di Fisica, Sapienza
  Universit\`a di Roma, Piazzale Aldo Moro 2, Roma 00185, Italy}
\affiliation{Istituto Nazionale di Fisica Nucleare, Sezione di Roma 1,
  Roma 00185, Italy}
\author{Giovanni Jona-Lasinio} \affiliation{Dipartimento di Fisica,
  Sapienza Universit\`a di Roma, Piazzale Aldo Moro 2, Roma 00185,
  Italy} \affiliation{Istituto Nazionale di Fisica Nucleare, Sezione
  di Roma 1, Roma 00185, Italy}

\date{\today}

\begin{abstract}
  According to quantum mechanics chiral molecules, that is molecules
  that rotate the polarization of light, should not exist.  The
  simplest molecules which can be chiral have four or more atoms with
  two arrangements of minimal potential energy that are equivalent up
  to a parity operation.  Chiral molecules correspond to states
  localized in one potential energy minimum and can not be stationary
  states of the Schr\"odinger equation. A possible solution of the
  paradox can be founded on the idea of spontaneous symmetry breaking.
  This idea was behind work we did previously involving a localization
  phase transition: at low pressure the molecules are delocalized
  between the two minima of the potential energy while at higher
  pressure they become localized in one minimum due to the
  intermolecular dipole-dipole interactions.  Evidence for such a
  transition is provided by measurements of the inversion spectrum of
  ammonia and deuterated ammonia at different pressures.  A previously
  proposed model gives a satisfactory account of the empirical results
  without free parameters.  In this paper, we extend this model to gas
  mixtures.  We find that also in these systems a phase transition
  takes place at a critical pressure which depends on the composition
  of the mixture. Moreover, we derive formulas giving the dependence
  of the inversion frequencies on the pressure.  These predictions are
  susceptible to experimental test.
\end{abstract}
\pacs{34.10.+x; 03.65.Xp; 34.20.Gj}
\maketitle

\section{Introduction}

The existence of chiral molecules, that is molecules that rotate the
polarization of light, has so far no universally accepted
explanation. Hund was the first to point out in 1927 that such
molecules according to quantum mechanics should not exist as the
corresponding Hamiltonian is invariant under mirror reflection
\cite{Hund}.  This problem became known as Hund's paradox and was
discussed in the physico-chemical literature for decades.  A possible
solution of the paradox can be founded on the idea of spontaneous
symmetry breaking (SSB) once we take into account that the experiments
deal with a large number of molecules and that their interactions are
sufficient to keep them in the chiral state.  This idea was behind
work done previously
\cite{Anderson,jonaclaverie,jonaclaverie2,Wightman,gm,jptprl,Jona}
where the explanation of the existence was based on a localization
phenomenon.

\begin{figure}[t]
  \includegraphics[width=0.33\columnwidth,clip]{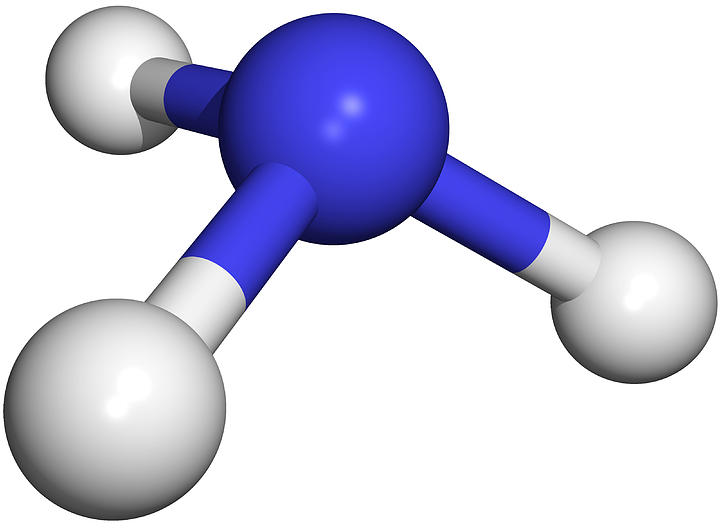}
  \qquad\qquad
  \includegraphics[width=0.4\columnwidth,clip]{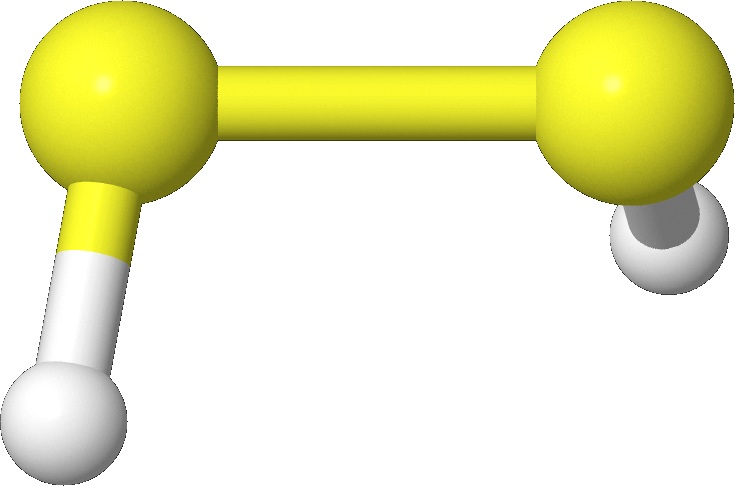}
  \centering
  \caption{(Color online) Molecular structures of ammonia
    ($\mathrm{NH_3}$, left) and deuterated disulfane
    ($\mathrm{D_2S_2}$, right) in one of their two localized states.}
  \label{NH3D2S2}
\end{figure}

Pyramidal molecules, i.e., molecules of the kind $XY_3$ such as
ammonia $\mathrm{NH}_3$, phosphine $\mathrm{PH}_3$, arsine
$\mathrm{AsH}_3$, are potentially chiral.  Suppose that we replace two
of the hydrogens with different atoms such as deuterium and tritium:
we obtain a molecule of the form $XYWZ$. This is called an enantiomer,
that is a molecule whose mirror image can not be superimposed to the
original one. These molecules are optically active in states where the
atom X is localized on one side of the $YWZ$ plane.  In the following
we shall refer to potentially chiral molecules such as ammonia as
pre-chiral molecules.  Another example of pre-chiral molecule, not
pyramidal, is deuterated disulfane $\mathrm{D}_2\mathrm{S}_2$ (see
Fig.~\ref{NH3D2S2}).

According to quantum mechanics chiral molecules should not exist as
stable stationary states.  Consider a pyramidal molecule $XY_3$. The
two possible positions of the $X$ atom with respect to the plane of
the $Y$ atoms are separated by a potential barrier and can be
connected via tunneling.  This gives rise to stationary wave functions
delocalized over the two minima of the potential and of definite
parity. In particular, the ground state is expected to be even under
parity. Tunneling induces a doublet structure of the energy levels.

On the other hand, the existence of chiral molecules can be
interpreted as a phase transition.  In fact, isolated molecules do not
exist in nature and the effect of the environing molecules must be
taken into account.  This interpretation underlies a simple mean-field
model that we have developed \cite{jptprl} to describe the transition
of a gas of $\mathrm{NH}_3$ molecules from a nonpolar phase to a polar
one through a localization phenomenon which gives rise to the
appearance of an electric dipole moment.  Even if ammonia molecules
are only pre-chiral, the mechanism, as emphasized in \cite{jptprl},
provides the key to understand the origin of chirality.

A quantitative discussion of the collective effects induced by
coupling a molecule to the environment constituted by the other
molecules of the gas was made in \cite{jonaclaverie}. In this work it
was shown that, due to the instability of tunneling under weak
perturbations, the order of magnitude of the molecular dipole-dipole
interaction may account for localized ground states.  This suggested
that a transition to localized states should happen when the
interaction among the molecules is increased.

Evidence for such a transition was provided by measurements of the
dependence of the doublet frequency under increasing pressure: the
frequency vanishes for a critical pressure $P_\mathrm{cr}$ different
for $\mathrm{NH}_3$ and $\mathrm{ND}_3$. The measurements were taken
at the end of the 1940s and beginning of the 1950s
\cite{Bleaney-Loubster.1948,Bleaney-Loubster.1950,Birnbaum-Maryott.1953b}
but no quantitative theoretical explanation was given for 50
years. Our model \cite{jptprl} gives a satisfactory account of the
empirical results.  A remarkable feature of the model is that there
are no free parameters. In particular, it describes quantitatively the
shift to zero-frequency of the inversion line of $\mathrm{NH}_3$ and
$\mathrm{ND}_3$ on increasing the pressure.

In this paper, we extend our model to gas mixtures.  This case may be
of interest, among other things, for the interpretation of the
astronomical data such as those from Galileo spacecraft \cite{HSK}
which measured the absorption spectrum of $\mathrm{NH}_3$ in the
Jovian atmosphere.

The main results of this work are as follows.

(i) Also in the case of mixtures there is a quantum phase transition
between a delocalized (or achiral, or nonpolar) phase and a localized
(or chiral, or polar) phase. At fixed temperature, the crossover
between the two phases is determined by a properly defined critical
pressure $P_\mathrm{cr}$.

(ii) We derive formulas expressing the critical pressure of the
mixture in terms of the doublet splittings of the isolated chiral
molecules and the fractions of the constituents.  For a binary
mixture with fractions $x_1$ and $x_2$ of the two species, we
have
\begin{align*}
  \frac{1}{P_\mathrm{cr}} = \sum_{i=1}^{2} x_i
  \frac{1}{P^{(i)}_\mathrm{cr}}.
\end{align*}
Supposing that species 1 is chiral, $P^{(1)}_\mathrm{cr}$ is the
critical pressure of the pure unmixed species, namely,
$P^{(1)}_\mathrm{cr} = \Delta E_1/(2\gamma_{11})$, with $\Delta E_1$
being the rarefied-gas doublet splitting and $\gamma_{11}$ given in
terms of the temperature and other microscopic parameters, i.e., the
electric dipole moment and the collision diameter of molecules 1 [see
Eq.~(\ref{gammaij})].  If species 2 is non polar, then
$P^{(2)}_\mathrm{cr} = \Delta E_1/(2\gamma_{12})$ with $\gamma_{12}$
expressing the polarization of molecules of species 2 by the chiral
molecules of species 1 [see Eq.~(\ref{G12})].  If species 2 is chiral,
then $P^{(2)}_\mathrm{cr} = \Delta E_2/(2\gamma_{22})$ with
$\gamma_{22}$ expressing the self-consistent interaction among the
molecules of species 2 [see Eq.~(\ref{gammaij})].  In both cases, the
inverse critical pressure of the mixture is the fraction-weighted
average of the inverse critical pressures of its components. The
result is easily generalized to mixtures made up of an arbitrary
number of components.

(iii) We derive formulas giving the dependence of the inversion
frequencies on the pressure [see Eqs.~(\ref{unifunmix}) and Sec.
\ref{inversionfrequencies.chimix}]. These results are susceptible to
experimental test.

The paper is structured as follows.  In the next section we briefly
recall the mean-field model for a single gas of molecules exhibiting
inversion doubling such as ammonia.  In particular, we discuss the
interaction mechanism and the strength of the interaction leading to a
localization quantum phase transition with the disappearance of the
inversion line.  In Sec. III, we describe the well known reaction
field mechanism which provides exactly the same interaction
strength. The reason for arguing in terms of the reaction field is its
simplicity especially in view of its extension to gas mixtures.  In
Sec. IV, we consider mixtures of a chiral gas with non polar
molecules, such as, for instance,
$\mathrm{NH}_3$--$\mathrm{H}_2$--$\mathrm{He}$.  In Sec. V, we
consider mixtures of chiral gases.  For convenience of the reader, we
have added three appendixes.  They give details on the model
\cite{jptprl} and the comparison with the experiments
\cite{Bleaney-Loubster.1948,Bleaney-Loubster.1950,Birnbaum-Maryott.1953b}.

\section{Chiral molecules as a case of spontaneous symmetry breaking}
\label{model}

We model a gas of pre-chiral, e.g., pyramidal, molecules as a set of
two-level quantum systems, that mimic the inversion degree of freedom
of an isolated molecule, mutually interacting via the dipole-dipole
electric force.

The Hamiltonian for the single isolated molecule is assumed of the
form $-\sigma^x \Delta E/2$, where $\Delta E$ is the inversion energy
splitting measured in a rarefied gas of the considered species and
\begin{align*}
  \sigma^x = \left( \begin{array}{cc} 0&1\\1&0 \end {array} \right)
\end{align*}
is the Pauli matrix with symmetric and antisymmetric delocalized
tunneling eigenstates $\varphi_+$ and $\varphi_-$:
\begin{align*}
  \varphi_+ = \frac{1}{\sqrt 2} \left( \begin{array}{c} 1 \\ 1 \end
    {array} \right), \qquad \varphi_- = \frac{1}{\sqrt 2}
  \left( \begin{array}{c} \phantom{-}1 \\ -1 \end {array} \right).
\end{align*}
Since the rotational degrees of freedom of an isolated molecule are
faster than the inversion ones, on the time scales of the inversion
dynamics, namely, $\hbar/\Delta E$, the molecules in a gas feel an
attraction arising from the angle averaging of the effective
dipole-dipole interaction at the actual temperature of the gas
\cite{Keesom}.  The localizing effect of the dipole-dipole interaction
between two molecules $i$ and $j$ can be represented by an interaction
term of the form $-\mathfrak{g}_{ij} \sigma^z_i \otimes \sigma^z_j$,
with $\mathfrak{g}_{ij}>0$, where
\begin{align*}
  \sigma^z = \left( \begin{array}{cc} 1&\phantom{-}0\\0&-1 \end
    {array} \right)
\end{align*} 
is the Pauli matrix with left and right localized eigenstates
$\varphi_L$ and $\varphi_R$
\begin{align*}
  \varphi_L = \left( \begin{array}{c} 1 \\ 0 \end {array} \right),
  \qquad \varphi_R = \left( \begin{array}{c} 0 \\ 1 \end {array}
  \right).
\end{align*}
The Hamiltonian for $N$ interacting molecules then reads as
\begin{align}
  H = -\frac{ \Delta E}{2} \sum_{i=1}^N \sigma^x_i - \sum_{i=1}^{N}
  \sum_{j=i+1} ^{N} \mathfrak{g}_{ij} \, \sigma^z_i \otimes
  \sigma^z_j.
  \label{HN}
\end{align}

For a gas of moderate density, we approximate the behavior of the $N
\gg 1$ molecules with the mean-field Hamiltonian
\begin{align}
  h(\psi)=-\frac{\Delta E}{2}\sigma^x- \frac{G}{N}\scp{\psi}{\sigma^z
    \psi} \sigma^z,
  \label{acca}
\end{align}
where $\psi$ is the mean-field molecular state to be determined
self-consistently by solving the nonlinear eigenvalue problem
associated with Eq.~(\ref{acca}) and the normalization condition
$\scp{\psi}{\psi}=N$.  The scalar product between two Pauli spinors
$\psi$ and $\phi$ is defined in terms of their two components in the
standard way
\begin{align*}
  \scp{\psi}{\phi} = \overline{\psi}_1\phi_1 +
  \overline{\psi}_2\phi_2.
\end{align*}

The parameter $G$ accounts for the effective dipole interaction energy
of a single molecule with the rest of the gas.  It must be identified
with a sum over all possible molecular distances and all possible
dipole orientations calculated assuming that, concerning the
translational, vibrational, and rotational degrees of freedom, the $N$
molecules behave as an ideal gas at thermal equilibrium at temperature
$T$. This assumption relies on a sharp separation (decoupling) between
these degrees of freedom and the inversion motion
\cite{Townes-Schawlow}.

We now discuss the calculation of $G$ following Keesom \cite{Keesom}.
An alternative derivation of $G$ will be shown later based on the so
called reaction field mechanism.

Let us define an effective interaction potential $V(r)$ between two
electric dipoles at distance $r$ via the formula
\begin{align*}
  e^{-V(r)/k_BT} = \langle e^{-V(r,\pmb{\mu}_1,\pmb{\mu}_2)/k_BT}
  \rangle ,
\end{align*}
where
\begin{align*}
  V(r,\pmb{\mu}_1,\pmb{\mu}_2) = \frac{ \pmb{\mu}_1 \cdot \pmb{\mu}_2
    - 3 (\pmb{\mu}_1 \cdot \pmb{r})(\pmb{r} \cdot \pmb{\mu}_2) r^{-2}}
  {4 \pi \varepsilon_0 r^3}
\end{align*}
is the standard dipole-dipole interaction potential and $\langle \dots
\rangle$ denotes the average over all possible orientations of the
vectors $\pmb{\mu}_1$ and $\pmb{\mu}_2$.  Assuming that $V(r) \ll k_B
T$ and observing that $\langle V(r,\pmb{\mu}_1,\pmb{\mu}_2)\rangle=0$,
we approximately evaluate the effective potential as
\begin{align*}
  V(r) &= -\frac{1}{2} \left\langle
    \left(\frac{V(r,\pmb{\mu}_1,\pmb{\mu}_2)}{k_BT}\right)^2
  \right\rangle k_BT
  \nonumber \\
  &= -\frac{\mu_1^2 \mu_2^2}{3(4\pi\varepsilon_0)^2 k_BT r^6}.
\end{align*}
In the case of a gas of molecules with dipole moments $\mu$, the
effective interaction energy of one molecule with the rest of the gas
amounts to $-G$, where
\begin{align}
  \label{G}
  G &= \int_d^\infty \rho(r) \frac{\mu^4}{3(4\pi\varepsilon_0)^2 k_BT
    r^6} 4\pi r^2 \d{r}
  \nonumber \\
  &= \frac{4\pi}{9}\frac{\mu^4 P}{(4 \pi \varepsilon_0 k_BT)^2 d^3}.
\end{align}
To get the above result we used $\rho(r)= P/k_BT$, the ideal gas
density at pressure $P$ and temperature $T$. Note that $d$ is the
minimal distance at which two molecules can be found, namely, the so
called molecular collision diameter.  At fixed temperature, the
effective interaction constant $G$ increases linearly with the gas
pressure $P$.

The nonlinear eigenvalue problem associated with Eq.~(\ref{acca}) is
solved in detail in Appendix \ref{stationarystateschiralgas}.  There
exists a critical value of the interaction strength,
$G_\mathrm{cr}=\Delta E/2$, such that for $G<G_\mathrm{cr}$ the
mean-field eigenstate with minimal energy is $\psi = \sqrt{N}
\varphi_+$, with delocalized molecules, whereas for $G>G_\mathrm{cr}$
there are two degenerate states of minimal energy. These are chiral
states transformed into each other by the parity operator $\sigma^x$
approaching the localized states $\varphi_L,\varphi_R$ for $G \gg
G_\mathrm{cr}$.  We thus have a quantum phase transition between a
delocalized (or achiral, or nonpolar) phase and a localized (or
chiral, or polar) phase.  In view of the dependence of $G$ on $P$, we
can define a critical pressure at which the phase transition takes
place
\begin{align}
  P_\mathrm{cr}= \frac{9}{8\pi} \frac{\Delta E d^3 (4 \pi
    \varepsilon_0 k_BT)^2}{\mu^4}.
  \label{pcr}
\end{align}
In Table \ref{pcr_tab} we report the values of $P_\mathrm{cr}$
calculated for different pre-chiral molecules.
\begin{table}
  \caption{\label{pcr_tab}
    Rarefied-gas energy splitting $\Delta E$, 
    electric dipole moment $\mu$, and collision diameter $d$ measured 
    for different pre-chiral molecules \cite{Townes-Schawlow,HCP}. 
    In the fourth column, we report the critical pressure $P_\mathrm{cr}$ 
    evaluated by Eq.~(\ref{pcr}) at $T=300$ K.}
  \begin{ruledtabular}
    \begin{tabular}{lcccc}
      &$\Delta E$ (cm$^{-1}$)&$\mu$ (Debye)&$d$ (\AA)&$P_\mathrm{cr}$ (atm)\\
      $\mathrm{NH_3}$&0.81&1.47&4.32&1.69\\
      $\mathrm{ND_3}$&0.053&1.47&4.32&0.11\\
      $\mathrm{D_2S_2}$&$10^{-9}$&1.56&5.97&$4.3\times 10^{-9}$\\
    \end{tabular}
  \end{ruledtabular}
\end{table}

In Appendix \ref{excitationsfromgroundstate} we show that in the
delocalized phase the inversion angular frequency of the interacting molecules
depends on the pressure as
\begin{align}
  \hbar \overline{\omega} = \Delta E \sqrt{1-\frac{P}{P_\mathrm{cr}}}.
  \label{unifun}
\end{align}
This formula is interesting as it expresses the ratio of two
microscopic quantities, $\hbar\overline{\omega}$ and $\Delta E$, as a
universal function of the ratio of the macroscopic variables $P$ and
$P_\mathrm{cr}$. Furthermore, it provides a very good representation
of some spectroscopic data, namely, the shift to zero frequency of the
inversion line of $\mathrm{NH}_3$ or $\mathrm{ND}_3$, as it appears
from Fig.~\ref{universal_line}, see Appendix \ref{inversion.line} for
more details.

\section{The reaction field mechanism}
\label{thereactionfieldmechanism}
The mean-field coupling constant $G$ introduced in Section \ref{model}
can be estimated, obtaining the same result, using a different
argument which can be easily generalized.  This is the reaction field
mechanism widely used in physics and chemistry
\cite{Onsager,Bottcher}.

Let us consider a spherical cavity of radius $a$ in a homogeneous
dielectric medium characterized by a relative dielectric constant
$\varepsilon_r$. An electric dipole $\pmb{\mu}$ placed at the center
of the cavity polarizes the dielectric medium inducing inside the
sphere a reaction field $\pmb{R}$ proportional to $\pmb{\mu}$:
\begin{align*}
  \pmb{R} = \frac{2(\varepsilon_r-1)}{2\varepsilon_r+1}
  \frac{\pmb{\mu}}{4 \pi \varepsilon_0 a^3}.
\end{align*}
As a result, the dipole acquires an energy
\begin{align}
  \label{dipole.energy}
  \mathscr{E} = - \frac{1}{2} \pmb{\mu}\cdot\pmb{R} = -
  \frac{\varepsilon_r-1}{2\varepsilon_r+1} \frac{\mu^2}{4 \pi
    \varepsilon_0 a^3}.
\end{align}
By using the Clausius--Mossotti relation
\begin{align*}
  \frac{\varepsilon_r-1}{\varepsilon_r+2} = \frac{1}{3} \rho
  \left(\alpha + \alpha^{\mathrm{D}} \right),
\end{align*}
where $\alpha$ is the molecular polarizability and
$\alpha^{\mathrm{D}} = \mu^2/(3 \varepsilon_0 k_B T)$ is the Debye
(orientation) polarizability, and observing that for a chiral gas
$\alpha^{\mathrm{D}} \gg \alpha$ (for instance, in the case of
$\mathrm{NH}_3$ we have $\alpha\simeq 2~\mbox{\AA}^3$ whereas
$\alpha^{\mathrm{D}}\simeq 217~\mbox{\AA}^3$ at $T=300~\mathrm{K}$),
we get
\begin{align*}
  \mathscr{E} = - \frac{4\pi}{9} \frac{\mu^4 P}{(4 \pi \varepsilon_0
    k_BT)^2a^3}.
\end{align*}
In all the above expressions we have $\varepsilon_r\simeq 1$ so that we
approximated $\varepsilon_r+2 \simeq 2\varepsilon_r +1 \simeq 3$.  

We exactly have $\mathscr{E}=-G$ provided $a=d$, where $d$ is the
molecular collision diameter introduced in Eq.~(\ref{G}) as the
minimum distance between two interacting molecules.  In fact, the size
of the spherical cavity is not arbitrary.  Microscopic arguments
\cite{Hynne-Bullough,Linder} show that $a$ must be identified with the
radius of an effective hard core, namely, the molecular collision
diameter $d$.

\section{Mixtures of a chiral gas with non polar molecules}
\label{mixture.chiral.nonpolar}
By using the reaction field approach to the coupling constant $G$, we
now generalize the mean-field model of Section \ref{model} to the case
of a mixture of chiral with non polar molecules.  Consider a gas
mixture of two species labeled 1 and 2.  In this case, the
Clausius-Mossotti relation reads as
\begin{align*}
  \frac{\varepsilon_r-1}{\varepsilon_r+2} = \frac{1}{3} \left( \rho_1
    \left(\alpha_1 + \alpha^{\mathrm{D}}_1 \right) + \rho_2
    \left(\alpha_2 + \alpha^{\mathrm{D}}_2 \right) \right).
\end{align*}
If species $1$ is pre-chiral, i.e., polar, $\alpha_1$ can be neglected
with respect to $\alpha^{\mathrm{D}}_1$ and if species $2$ is non
polar $\alpha^{\mathrm{D}}_2=0$.  The energy acquired by one molecule
of the polar species, having electric dipole moment $\mu_1$, then is
\begin{align}
  \label{E1mixture}
  \mathscr{E}_1 = - \frac{1}{3} \rho_1 \alpha^{\mathrm{D}}_1
  \frac{\mu_1^2}{4 \pi \varepsilon_0 d_{11}^3} - \frac{1}{3} \rho_2
  \alpha_2 \frac{\mu_1^2}{4 \pi \varepsilon_0 d_{12}^3},
\end{align}
where $d_{11}$ and $d_{12}$ are the diameters of the molecular
collisions $1$-$1$ and $1$-$2$, namely, $d_{11}=2r_1$ and $d_{12}=r_1+r_2$,
$r_1$ and $r_2$ being the hard sphere radii of the molecules 1 and 2.
If the density $\rho_1$ is such that $\rho_1 \alpha^{\mathrm{D}}_1 \ll
\rho_2 \alpha_2$, the above formula reduces to
\begin{align*}
  \mathscr{E}_1 = - \frac{1}{3} \rho_2 \alpha_2 \frac{\mu_1^2}{4 \pi
    \varepsilon_0 d_{12}^3}.
\end{align*}
This formula approximates the case of a single polar molecule immersed
in a non polar gas.

For a mixture of chiral and non polar molecules, the mean-field
molecular state of the chiral species 1, $\psi_1$, is determined
similarly to the case of a single chiral gas.  The only degree of
freedom of the non polar molecules is the deformation which, in turn,
is proportional to the electric dipole moment of the chiral molecules.
Therefore, we assume the following nonlinear Hamiltonian similar to
Eq.~(\ref{acca})
\begin{align}
  h_1(\psi_1)=-\frac{\Delta E_1}{2}\sigma_1^x
  -\frac{G_1}{N_1}\scp{\psi_1}{\sigma_1^z \psi_1} \sigma_1^z.
  \label{accamix}
\end{align}
The mean-field molecular state $\psi_1$ is normalized to the number of
molecules of the species 1, namely, $\scp{\psi_1}{\psi_1}=N_1$.  The
parameter $G_1$ accounts for the effective dipole interaction energy
of a single chiral molecule with the rest of the gas constituted by
polar and non polar species.  According to the reaction field
arguments, we have just to identify $G_1=-\mathscr{E}_1$, where
$\mathscr{E}_1$ is given by Eq.~(\ref{E1mixture}).  Thus we see that
$G_1$ is the sum of two contributions
\begin{align*}
  G_1=G_{11}+G_{12},
\end{align*}
where
\begin{align}
  \label{G11}
  G_{11} = \frac{4\pi}{9} \frac{\mu_1^4 P_1}{(4 \pi \varepsilon_0
    k_BT)^2d_{11}^3} = \gamma_{11} P_1
\end{align}
is the energy provided by the interaction with the chiral species and
\begin{align}
  \label{G12}
  G_{12} = \frac{1}{3} \frac{\alpha_2 \mu_1^2 P_2}{4\pi\varepsilon_0
    k_BT d_{12}^3} = \gamma_{12} P_2
\end{align}
the energy due to the presence of the non polar molecules.  In the
above expressions we used the ideal gas relations $\rho_1=P_1/k_BT$
and $\rho_2=P_2/k_BT$, $P_1$ and $P_2$ being the partial pressures of
the species 1 and 2.

The analysis of the mean-field Hamiltonian (\ref{accamix}) is
identical to the case of a single chiral gas. We have a localization
phase transition when $G_1$ becomes equal to ${G_1}_\mathrm{cr}=\Delta
E_1/2$.  The transition can be considered as a function of the total
pressure $P=P_1+P_2$ of the mixture and of the fractions of the
two species $x_1=P_1/P$ and $x_2=P_2/P$.  In this case, an important
prediction is that, instead of a unique critical pressure, we have a
critical line parametrized by $x_1$ (or $x_2=1-x_1$):
\begin{align}
  P_\mathrm{cr}(x_1) = 
  \frac{\Delta E_1}{2x_1 \gamma_{11} +2(1-x_1)\gamma_{12}},
  \label{pcrmix1}
\end{align}
where $\gamma_{11}=G_{11}/P_1$ and $\gamma_{12}=G_{12}/P_2$ 
with $G_{11}$ and $G_{12}$ as in Eqs.~(\ref{G11}) and (\ref{G12}).

The inversion frequency of the chiral species 1 can be evaluated as in
the case of a single chiral gas. The result is formally identical to
that in Eq.~(\ref{unifun}),
\begin{align}
  \hbar \overline{\omega}_1 = \Delta E_1
  \sqrt{1-\frac{P}{P_\mathrm{cr}(x_1)}},
  \label{unifunmix}
\end{align}
with the angular frequency $\overline{\omega}_1$ which is now a
function of the pressure $P$ of the mixture and of the fraction
$x_1$ of the chiral species.  However, for any fixed value of $x_1$,
the ratio $\hbar\overline{\omega}_1/\Delta E_1$ is still a universal
function of $P/P_\mathrm{cr}$ independent of the nature of species 1.

\begin{figure}
  \includegraphics[width=\columnwidth,clip]{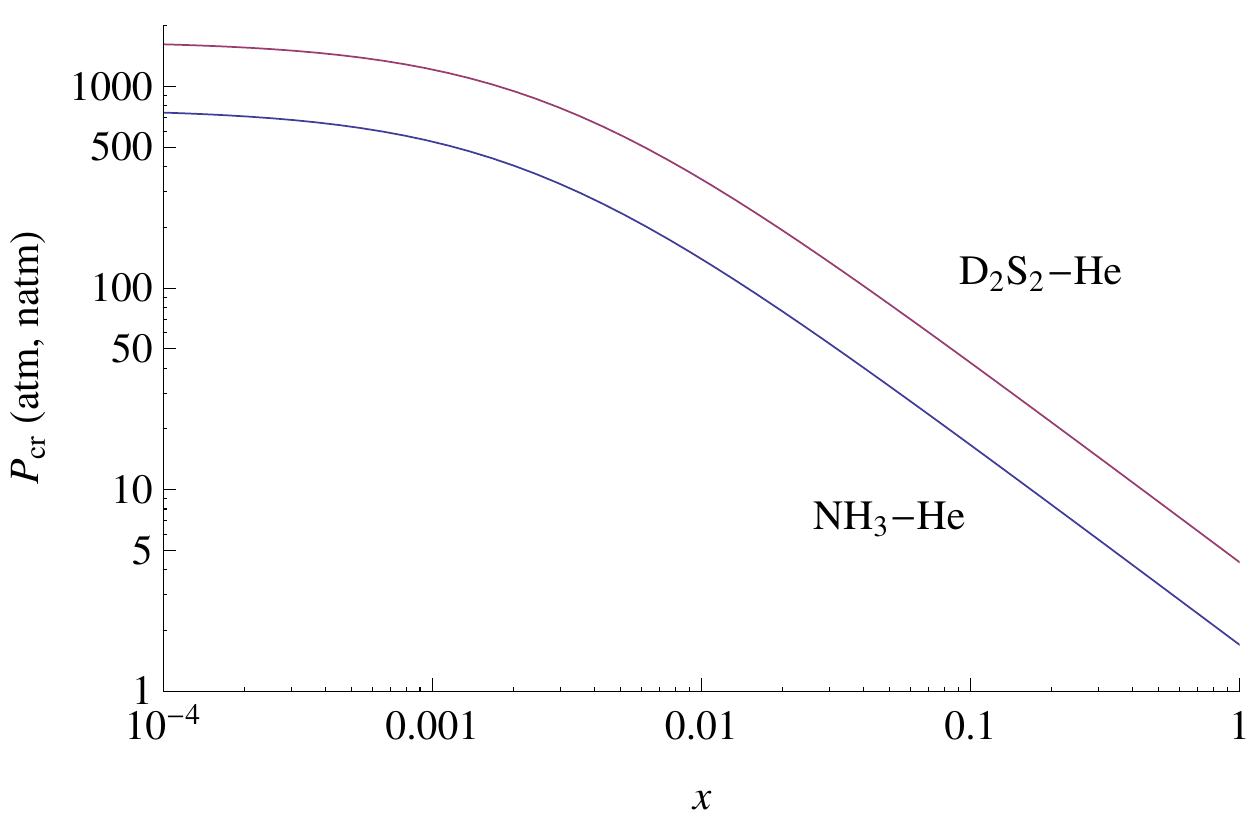}
  \centering
  \caption{(Color online) Critical pressure $P_\mathrm{cr}$ as a function of the
    chiral species fraction $x$ for NH$_3$--He (bottom) and
    D$_2$S$_2$--He (top) mixtures at $T=300$ K.  Note that pressure
    units are atm in the case of NH$_3$--He and natm in the case of
    D$_2$S$_2$--He.}
  \label{Pcr_mixture}
\end{figure}

In Fig.~\ref{Pcr_mixture}, we show the variation of the critical
pressure as a function of the fraction $x$ of the chiral species
for the mixtures NH$_3$--He and D$_2$S$_2$--He.  The value of
$P_\mathrm{cr}$ at $x=1$ coincides with that reported in Table
\ref{pcr_tab} for a pure chiral gas, whereas for $x\to 0$ the critical
pressure approaches a much higher value.  The ratio
\begin{align*}
  \frac{\lim_{x\to 0}P_\mathrm{cr}(x)}{P_\mathrm{cr}(1)} =
  \frac{4\pi}{3} ~\frac{\mu_1^2/(4 \pi \varepsilon_0d_{11}^3)}{k_BT}
  ~\frac{d_{12}^3}{\alpha_2}
\end{align*}
amounts to about 400 in the examples of Fig.~(\ref{Pcr_mixture}).

The above analysis is immediately generalized to a mixture containing
one chiral gas and an arbitrary number of non polar gases.  As an
example, for a ternary mixture (as before, species 1 is the chiral
one) we have a surface of critical pressures
\begin{align*}
  P_\mathrm{cr}(x_1,x_2) = 
  \frac{\Delta E_1}{2x_1 \gamma_{11} + 2x_2 \gamma_{12} + 
    2(1-x_1-x_2)\gamma_{13}}, 
\end{align*}
where $\gamma_{13}=G_{13}/P_3$ with $G_{13}$ given by an expression
analogous to Eq.~(\ref{G12}). This term expresses the polarization of
the molecules of species 3 by the chiral molecules of species 1.

The above results may be of interest, among other things, in the
interpretation of astronomical data such as those from Galileo spacecraft
\cite{HSK} which measured the absorption spectrum of NH$_3$ in the
Jovian atmosphere.  Note that, since the atmosphere of Jupiter is
mostly made of molecular hydrogen and helium in roughly solar
proportions whereas ammonia has a tiny fraction, the critical
pressure of the mixture is drastically different if compared with the
$1.69~\mathrm{atm}$ value of ammonia.  For fractions
$x_{\mathrm{NH}_3}\simeq 0.01$, $x_{\mathrm{H}_2}\simeq 0.86$ and
$x_\mathrm{He}\simeq 0.13$, we estimate $P_\mathrm{cr}$ ranging from
$15~\mathrm{atm}$ at $T=100~\mathrm{K}$ to $165~\mathrm{atm}$ at
$T=400~\mathrm{K}$.  This implies that no ammonia inversion line shift
can be observed in microwave absorption measurements performed in the
atmosphere of Jupiter.

\section{Mixtures of chiral gases}
\label{mixtures.chiral.gases}
It is possible to extend our mean-field approach to gas mixtures made
of several chiral species.  We start considering the simplest case of
two species.

In Sec. \ref{molecularstates.chimix}, we illustrate the model and
evaluate the stationary molecular states of the mixture.  In a first
reading, one can skip the discussion after Eq.~(\ref{psii}) and look
at the results summarized in Table~\ref{table_chimix}.  As in the case
of a single chiral gas, the mixture undergoes a localization phase
transition at a critical pressure $P_\mathrm{cr}$.  For
$0<P<P_\mathrm{cr}$, the lowest-energy molecular state of the mixture,
named $d_{++}$, corresponds to molecules of both species in a
delocalized symmetric configuration.  For $P>P_\mathrm{cr}$, new
minimal-energy molecular states appear with twofold degeneracy.  These
states, named $c_{LL}$ and $c_{RR}$, correspond to
molecules of both species in a chiral configuration of type $L$ or
$R$.  Equation (\ref{pcr_chimix_ave}) provides a simple formula for
$P_\mathrm{cr}$. The inverse critical pressure of the mixture is the
fraction-weighted average of the inverse critical pressures of
its components.

In Sec. \ref{inversionfrequencies.chimix}, we evaluate the
excitation energies of a two-species mixture by using the linear-response theory. The results are summarized by Eq.~(\ref{omegapm}).
We have two normal modes which at low pressure reduce to the inversion
of the component species.  By increasing the pressure, the energy of
both modes decreases, the smaller one vanishing at $P=P_\mathrm{cr}$,
the larger one approaching for $P\gg P_\mathrm{cr}$ the asymptotic
value given by Eq.~(\ref{omegapasympt}).  An example of this behavior
is shown in Figs.~\ref{inversionlineschimix1} and
\ref{inversionlineschimix2} in the case of a
$\mathrm{NH}_3$--$\mathrm{ND}_3$ mixture.

Finally, in Sec. \ref{consistency.chimix}, we discuss the
consistency of the developed theory by considering the absorption
coefficient and two limits in which it must reduce to that of a single
chiral gas, namely, the limit of zero fraction of one species and
the limit of equal species. We also discuss the possibility of
experimental tests for the considered $\mathrm{NH}_3$--$\mathrm{ND}_3$
mixture.

All the above results can be generalized to mixtures of several chiral
gases.

\subsection{Molecular states}
\label{molecularstates.chimix}
For a mixture of two chiral gases, the Clausius-Mossotti relation reads as
\begin{align*}
  \frac{\varepsilon_r-1}{\varepsilon_r+2} = \frac{1}{3} \left(
    \rho_1\alpha^{\mathrm{D}}_1 + \rho_2\alpha^{\mathrm{D}}_2 \right),
\end{align*}
where the Debye polarizations $\alpha_i^{\mathrm{D}} = \mu_i^2/(3
\varepsilon_0 k_B T)$, $i=1,2$, are given in terms of the molecular
electric-dipole moments, $\mu_1$ and $\mu_2$, of the two species.
According to Eq.~(\ref{dipole.energy}), due to mutual interactions the
molecules of species 1 and 2 acquire the energies
\begin{align*}
  \mathscr{E}_1 = -\frac{4\pi}{9}
  \frac{\mu_1^2\mu_1^2Px_1}{(4\pi\varepsilon_0 k_BT)^2 d_{11}^3}
  -\frac{4\pi}{9} \frac{\mu_1^2\mu_2^2Px_2}{(4\pi\varepsilon_0 k_BT)^2
    d_{11}^3},
\end{align*}
\begin{align*}
  \mathscr{E}_2 = -\frac{4\pi}{9}
  \frac{\mu_2^2\mu_2^2Px_2}{(4\pi\varepsilon_0 k_BT)^2 d_{22}^3}
  -\frac{4\pi}{9} \frac{\mu_2^2\mu_1^2Px_1}{(4\pi\varepsilon_0 k_BT)^2
    d_{22}^3}.
\end{align*}
As usual, we used the ideal gas relations $\rho_i=P_i/k_BT$, $i=1,2$,
and introduced the fractions $x_1=P_1/P$ and $x_2=P_2/P$, where
$P$ is the total pressure of the mixture.

We describe the inversion degrees of freedom of the two species by
mean-field molecular states $\psi_1,\psi_2$ normalized to the number
of molecules of the corresponding species, $\scp{\psi_1}{\psi_1}=N_1$,
$\scp{\psi_2}{\psi_2}=N_2$. Here, $\psi_1$ and $\psi_2$ are Pauli
spinors and the scalar product is defined as in Section \ref{model}.
By writing $\mathscr{E}_i = - \sum_{j=1}^{2} G_{i,j}$, $i=1,2$, where
\begin{align}
  G_{ij} = \frac{4\pi}{9} \frac{\mu_i^2\mu_j^2
    Px_j}{(4\pi\varepsilon_0 k_BT)^2 d_{ii}^3},
  \label{Gij}
\end{align}
and reasoning as in Sec. \ref{model}, we introduce the following
set of two nonlinear Hamiltonians:
\begin{subequations}
  \label{acca12}
  \begin{align}
    h_1(\psi_1,\psi_2) &= -\frac{\Delta E_1}{2}\sigma_1^x
    -\sum_{j=1}^{2} \frac{G_{1j}}{N_j}\scp{\psi_j}{\sigma_j^z \psi_j}
    \sigma_1^z,
    \\
    h_2(\psi_1,\psi_2) &= -\frac{\Delta E_2}{2}\sigma_2^x
    -\sum_{j=1}^{2} \frac{G_{2j}}{N_j}\scp{\psi_j}{\sigma_j^z \psi_j}
    \sigma_2^z.
  \end{align}
\end{subequations}
Pauli operators now have a label $i=1,2$ relative to the species they
refer to.  Stationary states are obtained by solving the eigenvalue problem 
\begin{subequations}
  \label{vectornlep}
  \begin{align}
    h_1(\psi_1,\psi_2) \psi_1 &= \lambda_1 \psi_1,  
    \\
    h_2(\psi_1,\psi_2) \psi_2 &= \lambda_2 \psi_2,
  \end{align}
\end{subequations}
where $\lambda_1,\lambda_2$ are the Lagrange multipliers associated with
the conservation of the particle number of species 1 and 2,
separately.  In fact, the molecules of the two species can not be
transformed into each another.

As discussed in the case of a single chiral gas (see Appendix
\ref{stationarystateschiralgas}) the molecules of the species 1 and 2
will occupy the eigenstate having the lowest energy
\begin{align}
  \mathcal{E}(\psi_1,\psi_2) =& -\sum_{i=1}^2 \frac{\Delta E_i}{2}
  \scp{\psi_i}{\sigma_i^x \psi_i} \nonumber \\ &
  -\sum_{i=1}^2\sum_{j=1}^2 \frac{G_{ij}}{2N_j}
  \scp{\psi_i}{\sigma_i^z \psi_i} \scp{\psi_j}{\sigma_j^z \psi_j}.
  \label{Epsi1psi2}
\end{align}
In virtue of the symmetry property $G_{12}/N_2=G_{21}/N_1$, the above 
energy is a constant of motion of the infinite-dimensional 
Hamiltonian system 
\begin{subequations}
  \begin{align*}
    i\hbar \frac{\d}{\d{t}} \psi_1 = h_1(\psi_1,\psi_2) \psi_1, 
    \\
    i\hbar \frac{\d}{\d{t}} \psi_2 = h_2(\psi_1,\psi_2) \psi_2. 
  \end{align*}
\end{subequations}
Notice, in fact,  that 
$\delta\mathcal{E}(\psi_1,\psi_2)/\delta \overline{\psi}_i =
h_i(\psi_1,\psi_2)\psi_i$.

To solve the eigenvalue problem (\ref{vectornlep}), we write
\begin{align}
  \psi_i = a_i \varphi_+^{(i)} + b_i \varphi_-^{(i)}, \qquad i=1,2.
  \label{psii}
\end{align}
The coefficients $a_i$ and $b_i$ can be chosen real and are normalized
to the number of molecules of each species, $a_i^2+b_i^2=N_i$.  Up to
an irrelevant global sign, we can also choose $a_1$ and $a_2$
non-negative.  By inserting the expressions (\ref{psii}) into
Eq.~(\ref{vectornlep}), we find that $a_1,b_1,a_2,b_2$ are the
solutions of the system of equations
\begin{align}
  \Delta E_i a_i b_i = \sum_{j=1}^{2} 2g_{ij} \left( a_i^2-b_i^2
  \right) a_j b_j, \qquad i=1,2,
  \label{aibi}
\end{align}
where $g_{ij}=G_{ij}/N_j$.  Once a solution of this system is found,
the corresponding energy can be evaluated as
\begin{align*}
  \mathcal{E} = - \sum_{i=1}^{2} \frac{\Delta E_i}{2} \left( a_i^2-b_i^2
  \right) - \sum_{i=1}^{2} \sum_{j=1}^{2} 2g_{ij} a_i b_i a_j b_j.
\end{align*}

It is convenient to discuss the solutions of Eq.~(\ref{aibi}) in
function of the parameter $P\in [0,\infty)$. In fact, the matrix
elements $g_{ij}$ are proportional to the pressure.  For any value of
$P$, Eq.~(\ref{aibi}) admits solutions such that $a_1b_1=0$ and
$a_2b_2=0$. There are four different
solutions satisfying this condition and the normalization rule
(see Table \ref{table_chimix}).
The solution
\begin{align}
  a_1=\sqrt{N_1},\quad b_1=0, \quad a_2=\sqrt{N_2},\quad b_2=0
  \label{delo}
\end{align}
is that with the lowest energy $\mathcal{E}=-N(x_1\Delta
E_1/2+x_2\Delta E_2/2) $.  We name this solution $d_{++}$
because it corresponds to molecules of both species in a delocalized
symmetric configuration.
\begin{widetext}
  \begin{table*}
    \caption{\label{table_chimix}
      Stationary molecular states for a mixture of two chiral gases
      with fractions $x_1=N_1/N$ and $x_2=N_2/N$, $N=N_1+N_2$ being 
      the total number of molecules.
      The states are classified in column 1 
      as delocalized ($d$) even ($+$) or odd ($-$) and chiral ($c$) left ($L$) 
      or right ($R$). 
      This classification results from the decomposition of the two-species
      molecular states in terms of the delocalized eigenfunctions of 
      $\sigma^x$ via the coefficients $a_1,b_1,a_2,b_2$ as in 
      Eq.~(\ref{psii}).
      The values of $a_1,b_1,a_2,b_2$ are found by solving Eq.~(\ref{aibi})
      with the normalization conditions $a_1^2+b_1^2=N_1$ and $a_2^2+b_2^2=N_2$.
      The chiral solutions exist only for $P>P_\mathrm{cr}$ and 
      are defined in terms of two parameters $q_1$ and $q_2$ 
      shown in Fig.~\ref{q1q2_chimix}.
      The energy per unit molecule 
      associated with each state [see Eq.~(\ref{Epsi1psi2})] 
      is reported in the last column.
      For $P<P_\mathrm{cr}$ 
      the solution of minimal energy is $d_{++}$ while for 
      $P>P_\mathrm{cr}$ there are two degenerate solutions of minimal energy,
      namely, $c_{LL}$ and $c_{RR}$.
      Other higher-energy states bifurcating at pressures larger 
      than $P_\mathrm{cr}$, see Fig.~\ref{states_chimix}, 
      are not reported in this table.}
    \begin{ruledtabular}
      \begin{tabular}{lccccc}
        &$a_1$&$b_1$&$a_2$&$b_2$&$\mathcal{E}/N$\\
        d$_{++}$&$\sqrt{N_1}$&0\phantom{$\sqrt{\frac{A_B}{C}}$}&$\sqrt{N_2}$&0&
        $-\frac{x_1\Delta E_1}{2}-\frac{x_2\Delta E_2}{2}$\\
        d$_{+-}$&$\sqrt{N_1}$&0\phantom{$\sqrt{\frac{A_B}{C}}$}&0&$\sqrt{N_2}$&
        $-\frac{x_1\Delta E_1}{2}+\frac{x_2\Delta E_2}{2}$\\
        d$_{-+}$&0\phantom{$\sqrt{\frac{A_B}{C}}$}&$\sqrt{N_1}$&$\sqrt{N_2}$&0&
        $+\frac{x_1\Delta E_1}{2}-\frac{x_2\Delta E_2}{2}$\\
        d$_{--}$&0\phantom{$\sqrt{\frac{A_B}{C}}$}&$\sqrt{N_1}$&0&$\sqrt{N_2}$&
        $+\frac{x_1\Delta E_1}{2}+\frac{x_2\Delta E_2}{2}$\\
        c$_{LL}$&
        $\sqrt{\frac{N_1}{2}(1+q_1)}$&$\sqrt{\frac{N_1}{2}(1-q_1)}$&
        $\sqrt{\frac{N_2}{2}(1+q_2)}$&$\sqrt{\frac{N_2}{2}(1-q_2)}$&
        $-\sum_{i}\frac{x_iq_i\Delta E_i}{2}
        -\sum_{i,j} \frac{x_ix_j\sqrt{1-q_i^2}\sqrt{1-q_j^2}\gamma_{ij} P}{2}$\\
        c$_{RR}$&
        $\sqrt{\frac{N_1}{2}(1+q_1)}$&$-\sqrt{\frac{N_1}{2}(1-q_1)}$&
        $\sqrt{\frac{N_2}{2}(1+q_2)}$&$-\sqrt{\frac{N_2}{2}(1-q_2)}$&
        $-\sum_{i}\frac{x_iq_i\Delta E_i}{2}
        -\sum_{i,j} \frac{x_ix_j\sqrt{1-q_i^2}\sqrt{1-q_j^2}\gamma_{ij} P}{2}$\\
      \end{tabular}
    \end{ruledtabular}
  \end{table*}
\end{widetext}

If $a_1b_1=0$ and $a_2b_2\neq 0$ or, vice versa,
$a_1b_1\neq 0$ and $a_2b_2=0$, Eq.~(\ref{aibi}) has no solutions.
In fact, if $a_1b_1=0$ then $a_1^2-b_1^2=\pm N_1$ and
Eq.~(\ref{aibi}) for $i=1$ gives $0=\pm 2g_{12}N_1 a_2b_2$ which
can not be solved for $a_2b_2\neq 0$.

Equation~(\ref{aibi}) admits solutions with
$a_1b_1\neq 0$ and $a_2b_2\neq 0$ if and only if
\begin{align}
  &\left[\Delta E_1-2g_{11}(a_1^2-b_1^2)\right] \left[\Delta
    E_2-2g_{22}(a_2^2-b_2^2)\right] \nonumber\\ &\qquad -4
  g_{12}g_{21}(a_1^2-b_1^2)(a_2^2-b_2^2) =0.
  \label{det=0}
\end{align}
In fact, think of Eq.~(\ref{aibi}) as a linear homogeneous system of
equations in the two unknowns $a_1b_1$ and $a_2b_2$. This system admits a
nontrivial solution $(a_1b_1,a_2b_2)\neq(0,0)$ if and only if its
determinant is 0, namely, Eq.~(\ref{det=0}). Any non trivial solution
is necessarily of the form $a_1b_1\neq 0$ and $a_2b_2\neq 0$ because
we have already established that Eq.~(\ref{aibi}) does not have
solutions with just one term between $a_1b_1$ and $a_2b_2$ equal to
0. 

If $(a_1,b_1,a_2,b_2)$ is a solution of Eq.~(\ref{aibi}), that is true also
for $(a_1,-b_1,a_2,-b_2)$ and these two solutions have the
same energy.  Therefore, each eigenstate obtained for $a_1b_1\neq 0$
and $a_2b_2 \neq 0$ has a twofold degeneracy.

Observing that $g_{11}g_{22}-g_{12}g_{21}=0$, Eq.~(\ref{det=0}) can be
rewritten as
\begin{align}
  &4g_{11}a_1^2\Delta E_2 + 4g_{22}a_2^2\Delta E_1 \nonumber \\
  &\qquad= 2g_{11}N_1\Delta E_2 + 2g_{22}N_2\Delta E_1 + \Delta E_1
  \Delta E_2.
  \label{det=0b}
\end{align}
The left-hand-side of Eq.~(\ref{det=0b}) is manifestly positive, therefore we
must conclude that also the right-hand-side is so.  In terms of $P$,
remember that $g_{ij}\propto P$, this condition implies that the
pressure has to be larger than the critical value
\begin{align}
  P_\mathrm{cr}=\left( \sum_{i=1}^2 \frac{2x_i \gamma_{ii}}{\Delta
      E_i} \right)^{-1},
  \label{pcr_chimix}
\end{align}
where we set
\begin{align}
  \gamma_{ij} = \frac{N}{P} g_{ij} = \frac{N}{N_jP} G_{ij}
  =\frac{4\pi}{9} \frac{\mu_i^2\mu_j^2}{(4\pi\varepsilon_0 k_BT)^2
    d_{ii}^3}.
  \label{gammaij}
\end{align}
For $x_1=1$ or $x_2=1$, Eq.~(\ref{pcr_chimix}) reduces to the critical
pressure (\ref{pcr}) for the single species 1 or 2.
\begin{figure}
  \includegraphics[width=\columnwidth,clip]{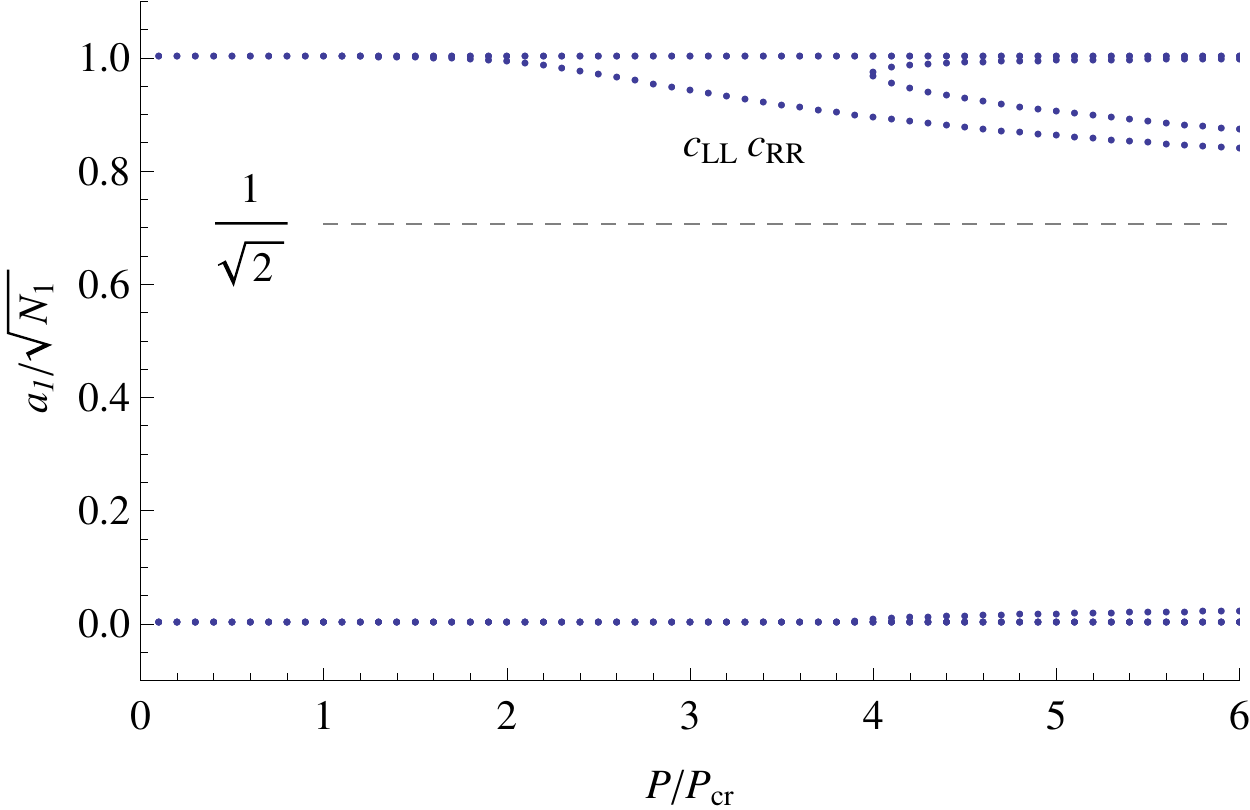}
  \includegraphics[width=\columnwidth,clip]{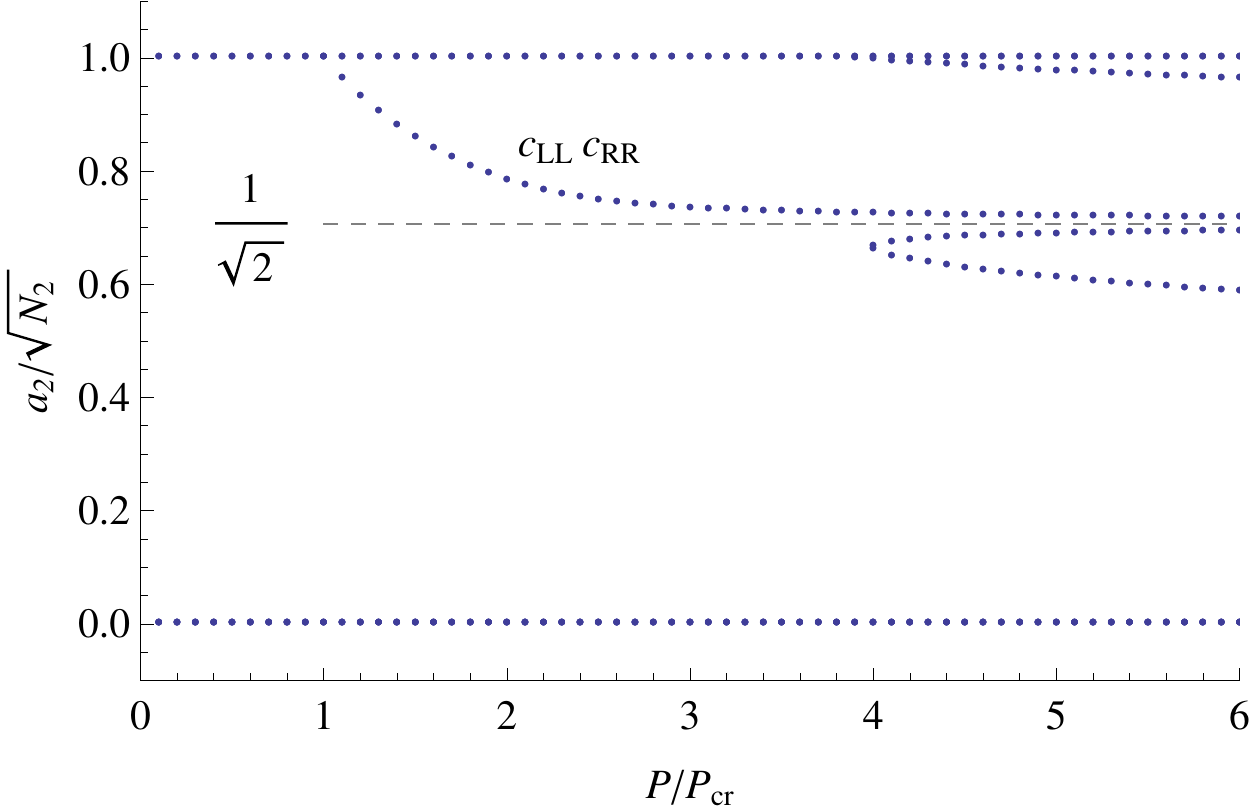}
  \centering
  \caption{(Color online) Coefficients $a_1/\sqrt{N_1}$ and
    $a_2/\sqrt{N_2}$ defining the molecular states $\psi_1,\psi_2$ of
    a two-species mixture as a function of $P/P_\mathrm{cr}$. The
    coefficients have been obtained by solving numerically
    Eq.~(\ref{aibi}) in the case of a $\mathrm{NH}_3$--$\mathrm{ND}_3$
    mixture ($\mathrm{ND}_3$ is species $i=2$) with $x_2=0.1$ at
    temperature $T=300~\mathrm{K}$.  The coefficients bifurcating at
    $P=P_\mathrm{cr}$ and tending for $P \gg P_\mathrm{cr}$ to
    $1/\sqrt{2}$ (dashed line) describe two degenerate molecular
    states, named $c_{LL}$ and $c_{RR}$, with
    molecules of both species in a chiral configuration of type $L$ or
    $R$. Notice further bifurcations appearing at $P \simeq 4
    P_\mathrm{cr}$, they correspond to stationary states of higher
    energy.}
  \label{states_chimix}
\end{figure}

In conclusion, similarly to the case of a single chiral gas, it exists
a critical pressure $P_\mathrm{cr}$ and a quantum phase transition
taking place at $P=P_\mathrm{cr}$.  For $0<P<P_\mathrm{cr}$, the
lowest-energy stationary state of the mixture is the delocalized
$d_{++}$ configuration given by Eq.~(\ref{delo}). At $P=P_\mathrm{cr}$
a bifurcation occurs and two new energy-degenerate stationary states
appear (see Fig.~\ref{states_chimix} for an example).  We call them
$c_{LL}$ and $c_{RR}$ because they correspond to
molecules of both species in a chiral configuration of type $L$ or
$R$ (see Table~\ref{table_chimix}), achieving a complete localization
for $P \gg P_\mathrm{cr}$. The value of $P_\mathrm{cr}$ is given by
Eq.~(\ref{pcr_chimix}).  This formula says that the inverse critical
pressure of the mixture is the fraction-weighted average of the
inverse critical pressures of its components,
\begin{align}
  \label{pcr_chimix_ave}
  \frac{1}{P_\mathrm{cr}} = \sum_{i=1}^{2} x_i
  \frac{1}{P^{(i)}_\mathrm{cr}}, \qquad P^{(i)}_\mathrm{cr} =
  \frac{\Delta E_i}{2\gamma_{ii}}.
\end{align}

\begin{figure}
  \includegraphics[width=\columnwidth,clip]{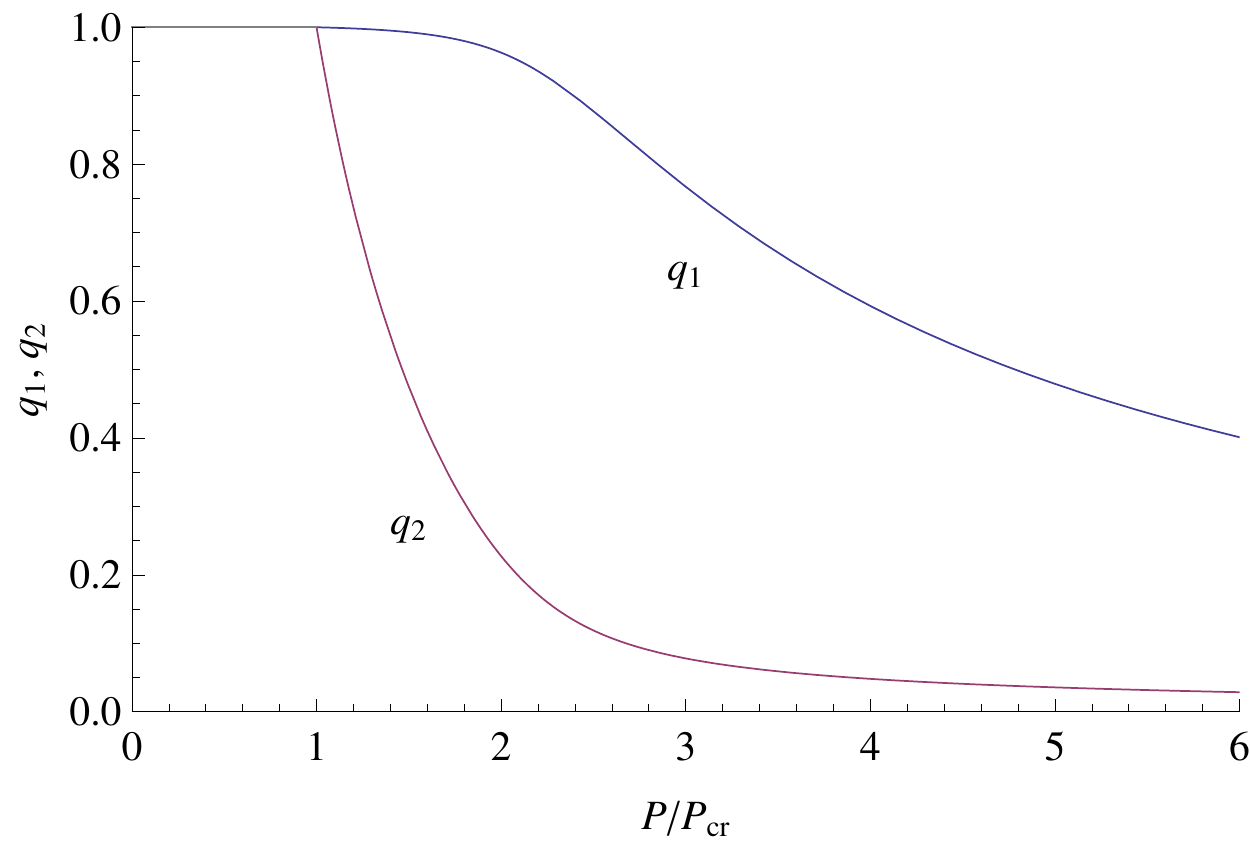}
  \centering
  \caption{(Color online) Parameters $q_1,q_2$ encoding the difference
    between the delocalized symmetric state $d_{++}$ and the
    chiral states $c_{LL}$ and $c_{RR}$ of
    Table~\ref{table_chimix} evaluated numerically for the same
    mixture of Fig.~\ref{states_chimix}.}
  \label{q1q2_chimix}
\end{figure}

The difference between the lowest-energy stationary solutions of the
two phases, namely, $d_{++}$ and $c_{LL}, c_{RR}$, can be entirely encoded in the parameters
\begin{align*}
  q_i = \left\{
    \begin{array}{ll}
      1, &\qquad P<P_\mathrm{cr},
      \\
      2a_i^2/N_i-1, &\qquad P>P_\mathrm{cr},       
    \end{array}
  \right. , \qquad i=1,2,
\end{align*}
(see Table~\ref{table_chimix}).  An example of the behavior of $q_1$ and
$q_2$ as a function of the pressure is shown in
Fig.~\ref{q1q2_chimix}.  Each parameter $q_i$ vanishes for $P\to
\infty$. More precisely, by using Eq.~(\ref{aibi}) it is simple to
show that at the leading order for $P$ large one has
\begin{align}
  \label{qias}
  q_i =
  \frac{\gamma_{ii}}{\sum_{j=1}^{2}\gamma_{ij}x_j}~
  \frac{P_\mathrm{cr}^{(i)}}{P},
\end{align}
where $P_\mathrm{cr}^{(i)}$ is the critical pressure of the $i$th
species as given by Eq.~(\ref{pcr_chimix_ave}).

\subsection{Excitation energies: Normal modes, inversion frequencies}
\label{inversionfrequencies.chimix}
As in the case of a single chiral species, we can determine the
excitation energies from the ground state of a binary mixture by
evaluating the linear response to an external time-dependent
perturbation representing an electro-magnetic radiation.  In detail,
we modify the Hamiltonians (\ref{acca12}) by adding for each species
$i=1,2$ a perturbation of the form $\mu_i F(t)\sigma_i^z$ with the
field $F(t)=F\cos(\omega t)$ sufficiently small so that effects of
order $O(F^2)$ are negligible.  The perturbation induces a
modified time evolution of the unperturbed ground state of the system
which we evaluate as follows.

For $P<P_\mathrm{cr}$, the ground state of the mixture at zero
external field is the symmetric delocalized state $d_{++}$
defined by the coefficients $(a_1,b_1,a_2,b_2)$ given by
Eq.~(\ref{delo}).  For $P>P_\mathrm{cr}$, the unperturbed mixture has
two lowest-energy degenerate stationary states, namely, the states
$c_{LL}$ and $c_{RR}$, defined by the two sets of
coefficients $(a_1,b_1,a_2,b_2)$ and $(a_1,-b_1,a_2,-b_2)$ given in
Table~\ref{table_chimix} in terms of the parameters $q_1,q_2$.  For a
racemic mixture, which is usually the case of concern, we may assume
that the unperturbed ground state of species $i=1,2$ in the localized
phase is represented by the \emph{stationary} density matrix obtained
by summing with equal weights the pure chiral states of types $L$ and
$R$
\begin{align}
  \varrho_i^{(0)} &= \frac{1}{2}
  \left(
    \begin{array}{c}
      a_i \\
      b_i
    \end{array}
  \right)
  \left(
    \begin{array}{cc}
      a_i & b_i
    \end{array}
  \right) + \frac{1}{2}
  \left(
    \begin{array}{c}
      a_i \\
      -b_i
    \end{array}
  \right)
  \left(
    \begin{array}{cc}
      a_i & -b_i
    \end{array}
  \right) \nonumber \\ &= \frac{N_i}{2} \left( \mathbb{1}_i +
    q_i\sigma_i^x \right).
  \label{rho0}
\end{align}
Since for $P<P_\mathrm{cr}$ we have $q_1=q_2=1$, Eq.~(\ref{rho0})
represents the ground-state density matrix of the species $i=1,2$ also in
the delocalized phase.  In conclusion, we can analyze simultaneously
both phases $P \gtrless P_\mathrm{cr}$ assuming that the unperturbed
lowest-energy stationary state is given by the density matrices
$\varrho_1^{(0)},\varrho_2^{(0)}$ of Eq.~(\ref{rho0}).  

Under the effect of the perturbation, the time evolution of the density 
matrices of the two species is governed by the equations
\begin{subequations}
  \label{vonNeumann}
  \begin{align}
    i\hbar \frac{\d}{\d{t}} \varrho_1(t) &=
    [h_1(\varrho_1(t),\varrho_2(t)) + \mu_1 F(t)\sigma_1^z,
    \varrho_1(t)],
    \\
    i\hbar \frac{\d}{\d{t}} \varrho_2(t) &=
    [h_2(\varrho_1(t),\varrho_2(t)) + \mu_2 F(t)\sigma_2^z,
    \varrho_2(t)],
  \end{align}
\end{subequations}
where the mean-field Hamiltonians $h_i(\varrho_1,\varrho_2)$, $i=1,2$, are
the same as Eq.~(\ref{acca12}) only rewritten in terms of the
density matrices $\varrho_1,\varrho_2$
\begin{align}
  h_i(\varrho_1,\varrho_2) = -\frac{\Delta E_i}{2}\sigma_i^x
  -\sum_{j=1}^{2} g_{ij} \trace{\sigma_j^z \varrho_j} \sigma_i^z.
  \label{accairho}
\end{align}
The density matrix of the $i$th species at time $t$ is written as
\begin{align}
  \varrho_i(t) = \varrho_i^{(0)} +
  x_i(t)\sigma_i^x+y_i(t)\sigma_i^y+z_i(t)\sigma_i^z.
  \label{rhoit}
\end{align}
Neglecting second order terms in the perturbation, the Bloch vector
$(x_i(t),y_i(t),z_i(t))$ is determined by the system of differential
equations
\begin{align*}
  \begin{array}{l}
    \hbar x'_i(t) = 0
    \\
    \hbar y'_i(t) = \Delta E_i z_i(t)
    -\sum_{j=1}^{2} 2N_iq_ig_{ij} z_j(t) + N_iq_i\mu_i F(t)
    \\
    \hbar z'_i(t) = -\Delta E_i y_i(t)
  \end{array} ,
\end{align*}
obtained by plugging the definition (\ref{rhoit}) into
Eq.~(\ref{vonNeumann}).  

We are interested in evaluating the electric dipole moments of each
species $i=1,2$, namely,
\begin{align*}
  p_i(t)=\mu_i \trace{\sigma_i^z \varrho_i(t)}=2\mu_iz_i(t).
\end{align*}
From the differential equations for the Bloch vector, we find that the
Fourier transforms $\tilde{p}_1(\omega),\tilde{p}_2(\omega)$ are given
by the system of algebraic equations
\begin{align}
  &\left( \hbar^2 \omega^2 - \Delta E_i^2 \right) \tilde{p}_i(\omega)
  + \sum_{j=1}^{2} 2N_iq_i\Delta E_i g_{ij} \tilde{p}_j(\omega)
  \nonumber\\ &\qquad = 2\mu_i^2N_iq_i \Delta E_i \tilde{F}(\omega),
  \qquad i=1,2,
  \label{lschimix}
\end{align}
where $\tilde{F}(\omega)$ is the Fourier transform of the applied
field.  The frequency-dependent electric susceptibility of the $i$th
species is finally evaluated as the ratio between the Fourier
transform of the polarization density of that species and the Fourier
transform of the applied field,
\begin{align*}
  \chi_i(\omega) = 
\frac{\tilde{p}_i(\omega)/(N/\rho)}
  {\varepsilon_0\tilde{F}(\omega)},
\end{align*}
where $N/\rho$ is the volume occupied by the mixture and
$\varepsilon_0$ the vacuum dielectric constant \cite{note.units}.  As
usual, the density of the mixture at pressure $P$ and temperature $T$
is assumed to be $\rho=P/k_B T$.  By determining $\tilde{p}_1(\omega)$
and $\tilde{p}_2(\omega)$ from Eq.~(\ref{lschimix}), we find that the
susceptibility of the mixture, $\chi=\chi_1+\chi_2$, is given by
\begin{align}
  \chi(\omega) =& \frac{\rho}{\varepsilon_0} \frac{C(\omega)}{D(\omega)},
  \label{chichimix}
\end{align}
where
\begin{align}
  C(\omega) =&\
    \xi_1\mu_1^2 \left[\hbar^2 \omega^2 - \Delta E_2^2 +
        \xi_2\left(\gamma_{22}-\gamma_{12}\right)P \right]  
    \nonumber \\ &+
    \xi_2\mu_2^2 \left[\hbar^2 \omega^2 - \Delta E_1^2 + 
        \xi_1\left(\gamma_{11}-\gamma_{21}\right)P \right],
  \label{Comega}
  \\
  D(\omega) =& \left( \hbar^2 \omega^2 - \Delta E_1^2 +
      \xi_1\gamma_{11}P \right) \nonumber\\ &\times 
\left(
    \hbar^2 \omega^2 - \Delta E_2^2 + \xi_2\gamma_{22}P 
  \right) - \xi_1\xi_2\gamma_{12}\gamma_{21}P^2,
  \label{Domega}
  \\
  \xi_i =&\ 2x_iq_i\Delta E_i, \qquad i=1,2.
  \label{xii}
\end{align}

The poles of the linear-response function (\ref{chichimix}) provide
the excitation energies from the ground state predicted by our
mean-field model \cite{Blaizot-Ripka}. It is worth to explore first two 
limit cases.

For noninteracting molecules, namely, $\gamma_{ij}=0$, $i,j=1,2$, we
have
\begin{align*}
  \chi(\omega) = \frac{\rho}{\varepsilon_0}\left(
    \frac{\xi_1\mu_1^2}{\hbar^2 \omega^2 - \Delta E_1^2}
    +\frac{\xi_2\mu_2^2}{\hbar^2 \omega^2 - \Delta E_2^2}\right).
\end{align*}
As expected, the response to an electromagnetic radiation
incident on the mixture would, in this case, provide
excitation energies $\Delta E_1$ and $\Delta E_2$ 
corresponding to the inversion of isolated molecules of
species 1 and 2.

For a gas with just one chiral species, let us say $x_2=0$,
Eq.~(\ref{chichimix}) reduces to
\begin{align*}
  \chi(\omega) = \frac{\rho\mu_1^2}{\varepsilon_0} \frac{2q_1\Delta
    E_1}{\hbar^2 \omega^2 - \left( \Delta E_1^2 - 2q_1\Delta
      E_1\gamma_{11}P \right) }.
\end{align*}
In this case we also have \cite{nota.gaspuro} $q_1=1$ for
$P<P_\mathrm{cr}$ and $q_1=\Delta E_1/(2 G_{11})$ for
$P>P_\mathrm{cr}$, where $G_{11}=\gamma_{11}P$, so that we recover the
result of Eq.~(\ref{linear.response.prl}).  The same conclusion is
obtained if we consider a single chiral gas as the mixture of two identical
species with arbitrary fractions $x_1$ and $x_2$ such that
$x_1+x_2=1$.

For a general mixture, no partial cancellation of zeros between
numerator and denominator of $\chi(\omega)$ takes place as in the
above two limit cases. The poles of the electric susceptibility
coincide with the zeros of $D(\omega)$.  By using the property
$\gamma_{11}\gamma_{22}-\gamma_{12}\gamma_{21}=0$, we find that the
bi-quadratic equation $D(\omega)=0$ has solutions with modulus
$\overline{\omega}_\pm$ given by
\begin{align}
  2\hbar^2\overline{\omega}_\pm^2 =& A_{12} \pm\sqrt{{A_{12}}^2 -
    4B_{12}},
  \label{omegapm}
\end{align}
where
\begin{align*}
  A_{12} &= \Delta E_1^2 + \Delta E_2^2 -
  (\xi_1\gamma_{11}+\xi_2\gamma_{22})P,
  \\
  B_{12} &= \Delta E_1^2\Delta E_2^2 - (\xi_1\gamma_{11}\Delta
  E_2^2+\xi_2\gamma_{22}\Delta E_1^2)P.
\end{align*}
Clearly, the excitation energies of the mixture
$\hbar\overline{\omega}_\pm$ are functions of the pressure.

At $P=0$, the normal modes represent the inversion of isolated
molecules of species 1 or 2, namely,
\begin{align*}
  \hbar\overline{\omega}_\pm(0)=\Delta E_{i_\pm},
\end{align*}
where $i_+=1$ and $i_-=2$ if $\Delta E_1>\Delta E_2$ and
$i_+=2$ and $i_-=1$ if $\Delta E_2>\Delta E_1$.

In the delocalized phase $0<P<P_\mathrm{cr}$, we have
\begin{align*}
  B_{12}(P) = \Delta E_1^2\Delta E_2^2 \left(1-P/P_\mathrm{cr} \right),
\end{align*}
in virtue of the fact that $q_1=q_2=1$.  For the same reason we also
have that $\xi_1$ and $\xi_2$ are constant in this phase so that
$A_{12}(P)$ decreases linearly with $P$.  As a result, we find that
$\hbar\overline{\omega}_\pm(P)$ decrease from their $P=0$ value
according to
\begin{align*}
  \hbar\overline{\omega}_\pm(P) \simeq 
  \Delta E_{i_\pm} \sqrt{1-\frac{P}{P_\mathrm{cr}^{(i_\pm)}}}.
\end{align*}
These approximated relations, which are merely the inversion frequency
laws for the pure species [see Eq.~(\ref{unifunapp})], are valid for
$P$ small.  The larger is the fraction of the species $i_\pm$,
the wider is the range of $P$ where the approximation is adequate.

At $P=P_\mathrm{cr}$, we have $B_{12}(P_\mathrm{cr})=0$ and, according
to Eq.~(\ref{omegapm}), the energy $\hbar\overline{\omega}_-$ vanishes
with an infinite slope. The energy of the other mode attains the
finite value $\hbar\overline{\omega}_+=\sqrt{A_{12}(P_\mathrm{cr})}$.

In the localized phase $P>P_\mathrm{cr}$, where $q_i=2a_i^2/N_i-1$,
$i=1,2$, with $a_1^2$ and $a_2^2$ satisfying Eq.~(\ref{det=0b}), we
have $B_{12}(P) = 0$ for any value of $P$.  It follows that
$\hbar\overline{\omega}_-=0$ throughout this phase.  The energy of the
other mode persists in decreasing and for $P\gg P_\mathrm{cr}$
approaches the asymptotic value
\begin{align}
  \hbar\overline{\omega}_+^\mathrm{as}= \sqrt{ \frac{\Delta
      E_1^2\gamma_{12}x_2}{\gamma_{11}x_1+\gamma_{12}x_2} +
    \frac{\Delta E_2^2\gamma_{21}x_1}{\gamma_{22}x_2+\gamma_{21}x_1}}.
  \label{omegapasympt}
\end{align}
This value corresponds to $\sqrt{A_{12}}$ evaluated with the leading
order expressions of $q_1,q_2$ as in Eq.~(\ref{qias}).

\begin{figure}
  \includegraphics[width=\columnwidth,clip]{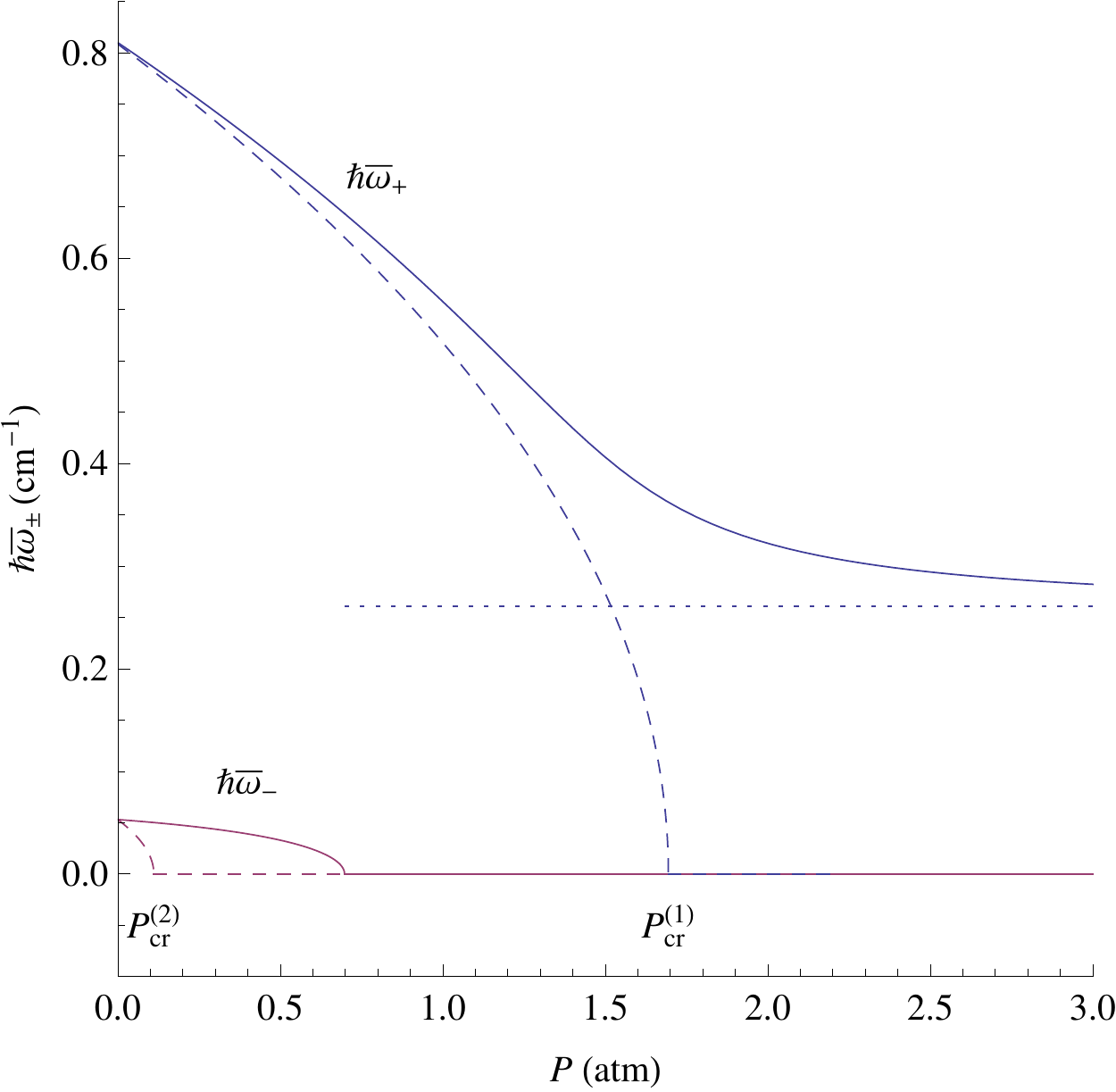}
  \centering
  \caption{(Color online) Energies $\hbar\overline{\omega}_\pm$ of the
    excitation normal modes as a function of pressure $P$ in the case
    of a $\mathrm{NH}_3$--$\mathrm{ND}_3$ mixture ($\mathrm{ND}_3$ is
    species $i=2$) with $x_2=0.1$ at $T=300~\mathrm{K}$. The
    horizontal dotted line is the asymptotic value
    $\hbar\overline{\omega}_+^\mathrm{as}$.  The dashed lines which
    vanish at $P_\mathrm{cr}^{(1)}=1.69~\mathrm{atm}$ and
    $P_\mathrm{cr}^{(2)}=0.11~\mathrm{atm}$ correspond to the
    inversion frequencies of pure species.  The critical pressure of
    the mixture, point where $\hbar\overline{\omega}_-$ vanishes, is
    $P_\mathrm{cr}=0.70~\mathrm{atm}$. Energy is measured in
    $\textrm{cm}^{-1}$.}
  \label{inversionlineschimix1}
\end{figure}

\begin{figure}
  \includegraphics[width=\columnwidth,clip]{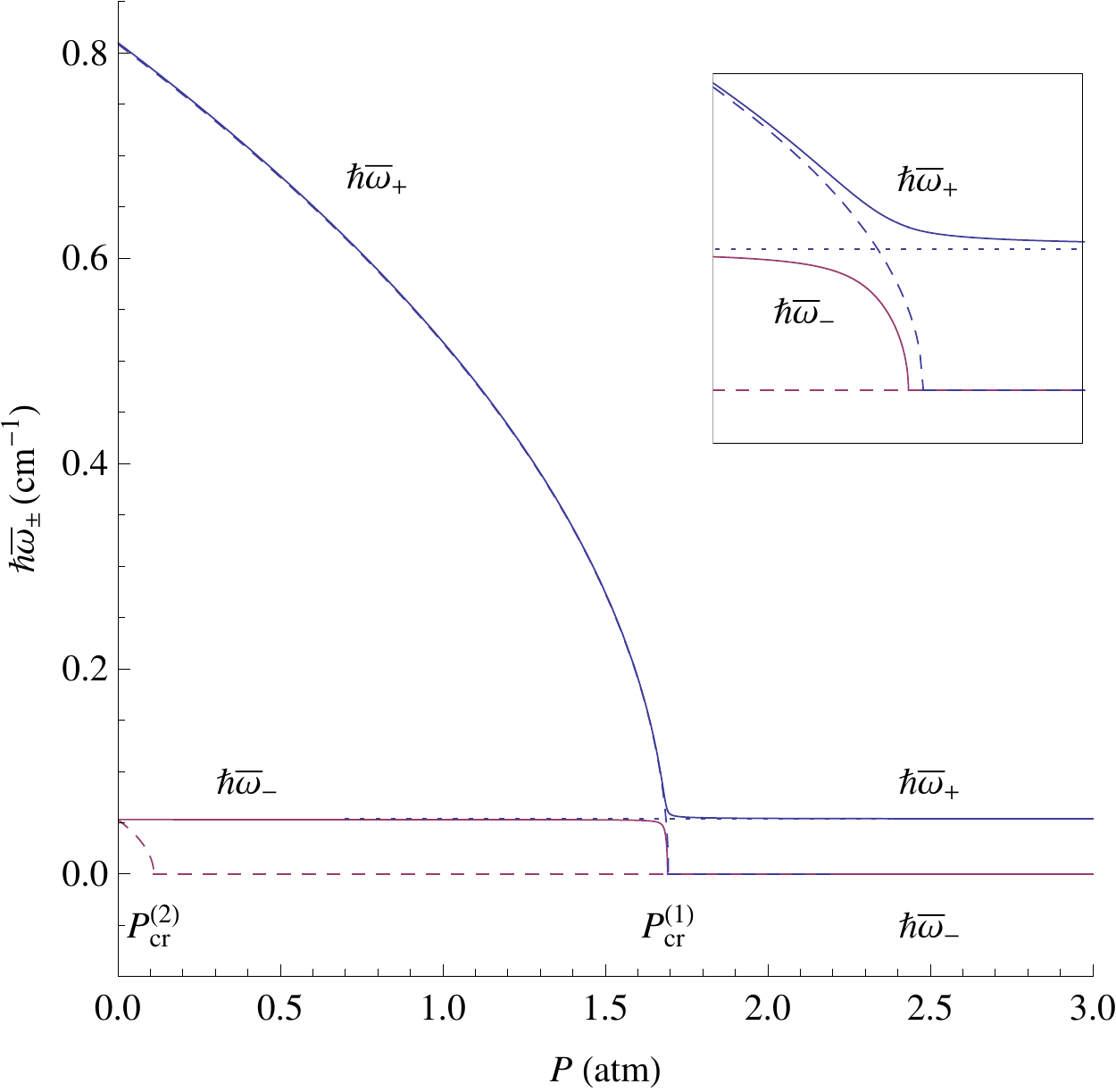}
  \centering
  \caption{(Color online) As in Fig.~\ref{inversionlineschimix1} for
    $x_2=0.0001$. For such small value of $x_2$ the critical pressure
    of the mixture is $P_\mathrm{cr} \simeq P_\mathrm{cr}^{(1)}$ and
    we have $\hbar\overline{\omega}_+^\mathrm{as} \simeq \Delta
    E_{i_-}$. Inset is a zoomed in view of the region $P\simeq
    P_\mathrm{cr}$.}
  \label{inversionlineschimix2}
\end{figure}

In Figs.~\ref{inversionlineschimix1} and \ref{inversionlineschimix2}
we show the excitation energies $\hbar\overline{\omega}_\pm$
calculated for a $\mathrm{NH}_3$--$\mathrm{ND}_3$ mixture with 10\%
and 0.01\% deuterated ammonia fraction, respectively.

\subsection{Consistency of the model: Absorption coefficient}
\label{consistency.chimix}
We have already discussed the consistency of our model by considering
two limiting cases in which it must recover the single gas case.  In
fact, we have observed that the susceptibility of a binary mixture
reduces to that of a single chiral gas in the limit of fraction
of one species tending to zero or in the limit in which the parameters
defining the two species become equal.  Here, we study in deeper
details this limiting behavior by taking into account the absorption
coefficient associated with the evaluated susceptibility.

According to the Beer-Lambert law, the intensity of a monochromatic
radiation of angular frequency $\omega$ propagating for a length
$\ell$ into the mixture is decreased by a factor $e^{-\kappa(\omega)
  \ell}$ due to the excitation of the normal modes.  Neglecting
deexcitation from these modes, i.e., spontaneous and stimulated
reemission of radiation, the absorption coefficient is
\begin{align}
  \kappa(\omega) = \frac{\omega}{c} \IM \chi(\omega),
\end{align}
with $c$ speed of light and $\chi(\omega)$ as in Eq.~(\ref{chichimix}).
The imaginary part of susceptibility is evaluated with the prescription 
$\omega \to \omega - i \eta$ where $\eta\to 0^+$. 
The result is 
\begin{align}
  \kappa(\omega) =& 
  I_+ \left[ \delta(\hbar\omega-\hbar\overline{\omega}_+) + 
    \delta(\hbar\omega+\hbar\overline{\omega}_+) \right]
  \\ \nonumber &+
  I_- \left[ \delta(\hbar\omega-\hbar\overline{\omega}_-) + 
    \delta(\hbar\omega+\hbar\overline{\omega}_-) \right],
\end{align}
where the intensities $I_\pm$ are 
\begin{align}
  I_\pm = \frac{\pi \rho}{2 \varepsilon_0 c \hbar}~
  \frac{C(\overline{\omega}_\pm)}
  {\hbar^2\overline{\omega}_\pm^2-\hbar^2\overline{\omega}_\mp^2},
  \label{Ipm}
\end{align}
with $C(\omega)$ given by Eq.~(\ref{Comega}).  These intensities
depend on the pressure and for $P\to 0$ at the leading order they are
\begin{align}
  \label{IpmPsmall}
  I_\pm(P) = x_{i_\pm} \frac{\pi \Delta E_{i_\pm} \mu_{i_\pm}^2}
  {\varepsilon_0 c \hbar k_BT} P.
\end{align}
We have assumed, as usual, $\rho=P/(k_B T)$ and used the property
$\hbar\overline{\omega}_\pm \to \Delta E_{i_\pm}$ for $P\to 0$. Thus, when
the density vanishes and the inter-molecular interactions can be
neglected, the absorption coefficient of the mixture is the sum of the
absorption coefficients of the two isolated species weighted by the
corresponding fractions.

Note that for non-interacting molecules, i.e., $\gamma_{ij}=0$,
$i,j=1,2$, the intensities $I_\pm$ are as in Eq.~(\ref{IpmPsmall}),
i.e.,  the absorption coefficient grows indefinitely with the pressure.
On the other hand, for interacting molecules we obtain that for $P\to
\infty$ the absorption coefficient approaches a constant value with
asymptotic intensities
\begin{align*}
  I_+^{\textrm{as}} =&\ 
  x_{1}
  \frac{\pi \Delta E_{1} \mu_{1}^2}{\varepsilon_0 c \hbar k_BT}
  P_\mathrm{cr}^{(1)}
  \frac{\gamma_{11}}{\gamma_{11}x_1+\gamma_{12}x_2}
  \left[ 1-
    \left(\frac{\Delta E_2}{\hbar\overline{\omega}_+^\mathrm{as}}\right)^2
  \right]
  \\
  &+ x_{2}
  \frac{\pi \Delta E_{2} \mu_{2}^2}{\varepsilon_0 c \hbar k_BT}
  P_\mathrm{cr}^{(2)}
  \frac{\gamma_{22}}{\gamma_{22}x_2+\gamma_{21}x_1}
  \left[ 1-
    \left(\frac{\Delta E_1}{\hbar\overline{\omega}_+^\mathrm{as}}\right)^2
  \right],
  \\
  I_-^{\textrm{as}} =&\ 
  x_{1}
  \frac{\pi \Delta E_{1} \mu_{1}^2}{\varepsilon_0 c \hbar k_BT}
  P_\mathrm{cr}^{(1)}
  \frac{\gamma_{11}}{\gamma_{11}x_1+\gamma_{12}x_2}
  \left(\frac{\Delta E_2}{\hbar\overline{\omega}_+^\mathrm{as}}\right)^2
  \\
  &+ x_{2}
  \frac{\pi \Delta E_{2} \mu_{2}^2}{\varepsilon_0 c \hbar k_BT}
  P_\mathrm{cr}^{(2)}
  \frac{\gamma_{22}}{\gamma_{22}x_2+\gamma_{21}x_1}
    \left(\frac{\Delta E_1}{\hbar\overline{\omega}_+^\mathrm{as}}\right)^2.
\end{align*}

To understand the limit of the absorption coefficient for, e.g., $x_2\to
0$, observe that when the fraction of species 2 vanishes we have
$P_\mathrm{cr} \to P_\mathrm{cr}^{(1)}$ and the limit frequencies are
\begin{align*}
  &\hbar \overline{\omega}_+ \to \left\{
    \begin{array}{ll}
      \Delta E_1\sqrt{1-P/P_\mathrm{cr}^{(1)}}, &\qquad P<P_\mathrm{cr}^{(1)}
      \\
      \Delta E_2, &\qquad P>P_\mathrm{cr}^{(1)} 
    \end{array}
  \right.,
  \\
  &\hbar \overline{\omega}_- \to \left\{
    \begin{array}{ll}
      \Delta E_2, &\qquad P<P_\mathrm{cr}^{(1)}
      \\
      0, &\qquad P>P_\mathrm{cr}^{(1)}
    \end{array}
  \right.,
\end{align*}
if $\Delta E_1 > \Delta E_2$, whereas for $\Delta E_2 > \Delta E_1$
we have
\begin{align*}
  &\hbar \overline{\omega}_+ \to \Delta E_2, 
  \\
  &\hbar \overline{\omega}_- \to \left\{
    \begin{array}{ll}
      \Delta E_1\sqrt{1-P/P_\mathrm{cr}^{(1)}}, &\qquad P<P_\mathrm{cr}^{(1)}
      \\
      0, &\qquad P>P_\mathrm{cr}^{(1)}
    \end{array}
  \right..
\end{align*}
From Eq.~(\ref{Ipm}) it follows that 
\begin{align*}
  &I_+ \to \frac{\pi \Delta E_1 \mu_1^2}{\varepsilon_0 c \hbar k_B T} 
  \left\{
    \begin{array}{ll}
      P, &\qquad P<P_\mathrm{cr}^{(1)}
      \\
      0, &\qquad P>P_\mathrm{cr}^{(1)}
    \end{array}
  \right.,
  \\
  &I_- \to \frac{\pi \Delta E_1 \mu_1^2}{\varepsilon_0 c \hbar k_B T} 
  \left\{
    \begin{array}{ll}
      0, &\qquad P<P_\mathrm{cr}^{(1)}
      \\
      P_\mathrm{cr}^{(1)}, &\qquad P>P_\mathrm{cr}^{(1)}
    \end{array}
  \right.,
\end{align*}
if $\Delta E_1 > \Delta E_2$, whereas for $\Delta E_2 > \Delta E_1$
we have
\begin{align*}
  &I_+ \to 0,
  \\
  &I_- \to \frac{\pi \Delta E_1 \mu_1^2}{\varepsilon_0 c \hbar k_B T} 
  \left\{
    \begin{array}{ll}
      P &\qquad P<P_\mathrm{cr}^{(1)}
      \\
      P_\mathrm{cr}^{(1)} &\qquad P>P_\mathrm{cr}^{(1)}
    \end{array}
  \right..
\end{align*}
In both cases, we recover the intensity of the pure chiral gas 1,
increasing linearly with $P$ in the delocalized phase and locked at a
constant value in the localized one.  

In the limit $x_1 \to 0$ one has similar formulas which can be
obtained from the above ones by exchanging the indices $1
\leftrightarrow 2$.

\begin{figure}
  \includegraphics[width=\columnwidth,clip]{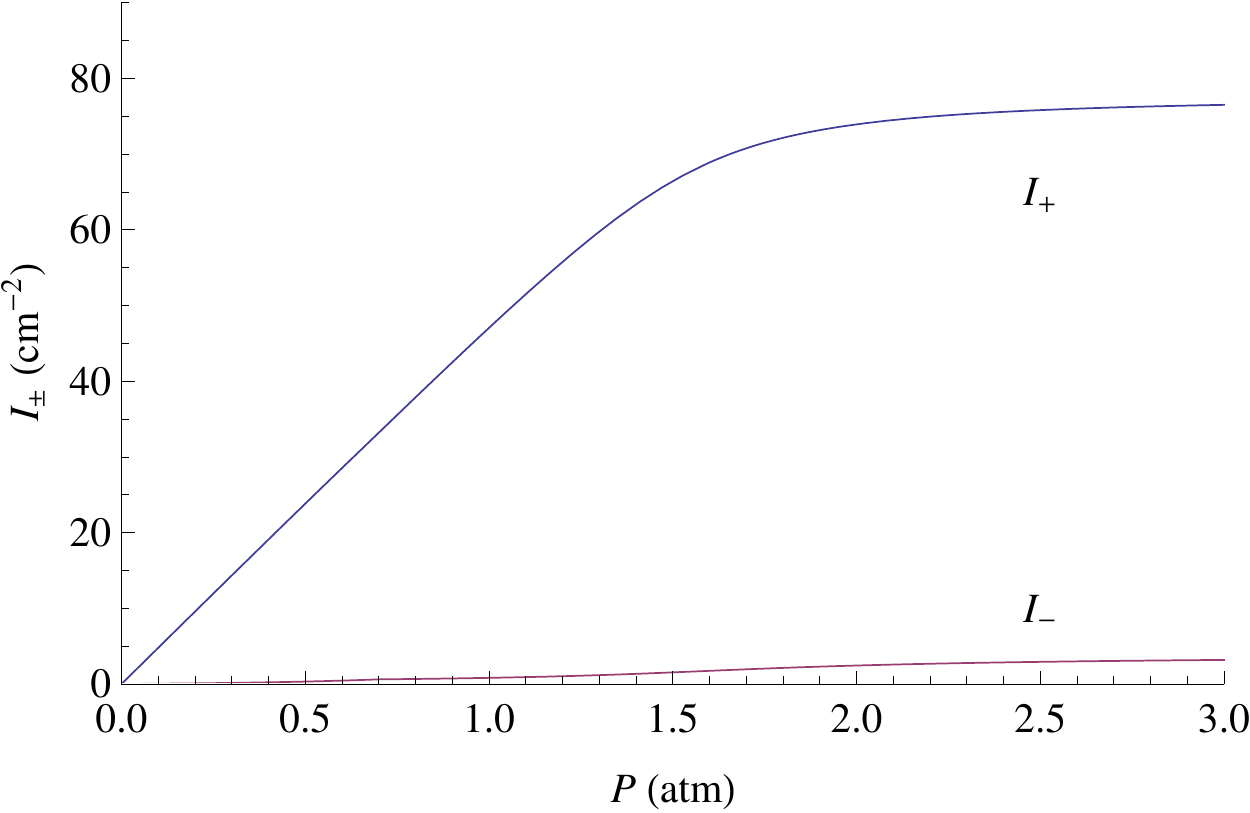}
  \centering
  \caption{(Color online) Intensities $I_\pm$ of the absorption
    coefficient as a function of pressure $P$ for the mixture of
    Fig.~\ref{inversionlineschimix1}. Energy is measured in
    $\textrm{cm}^{-1}$ and intensities, which have dimensions energy
    divided by length, are given in $\textrm{cm}^{-2}$.}
  \label{intensitieschimix1}
\end{figure}

\begin{figure}
  \includegraphics[width=\columnwidth,clip]{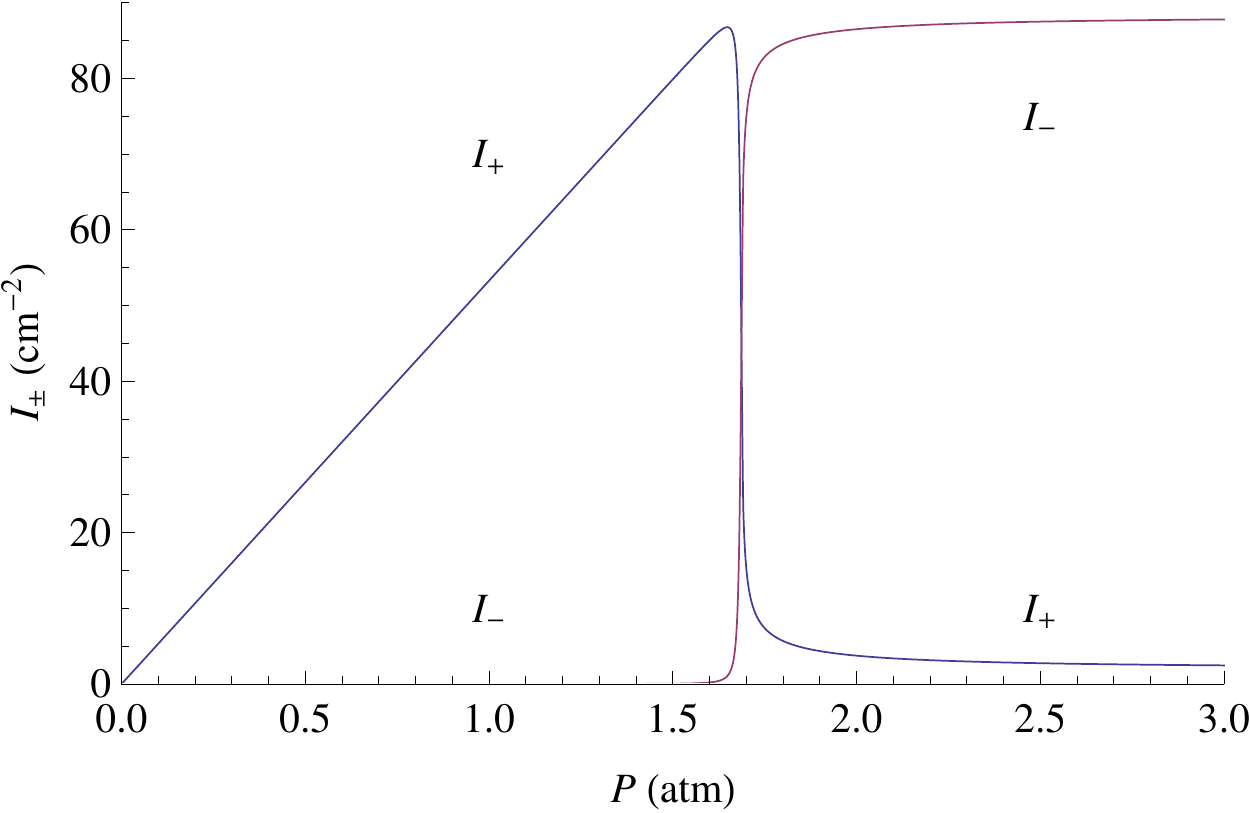}
  \centering
  \caption{(Color online) As in Fig.~\ref{intensitieschimix1} for the
    mixture of Fig.~\ref{inversionlineschimix2}.}
  \label{intensitieschimix2}
\end{figure}

An example of the behavior of the intensities $I_\pm$ as a function of
pressure $P$ is illustrated in Figs.  \ref{intensitieschimix1} and
\ref{intensitieschimix2} for the same $\mathrm{NH}_3$--$\mathrm{ND}_3$
mixtures considered in Figs.  \ref{inversionlineschimix1} and
\ref{inversionlineschimix2}, respectively.  In this case we have
$\Delta E_1 > \Delta E_2$ and the fraction $x_2$ decreases form
$0.1$, Figs. \ref{inversionlineschimix1} and \ref{intensitieschimix1},
to $0.0001$, Figs. \ref{inversionlineschimix2} and
\ref{intensitieschimix2}.  The dominant intensity, $I_+$ or $I_-$,
is mainly due to the more abundant species $i=1$.

We now discuss the limit in which $\Delta E_1, \Delta E_2 \to \Delta E$,
$\mu_1,\mu_2\to \mu$ and $\gamma_{i,j} \to \gamma$, $i,j=1,2$,  
whereas $x_1$ and $x_2$ are arbitrary with $x_1+x_2=1$.
In this case we have
\begin{align*}
  &\hbar \overline{\omega}_+ \to \Delta E,
  \\
  &\hbar \overline{\omega}_- \to 
  \left\{
    \begin{array}{ll}
      \Delta E \sqrt{1-P/P_\mathrm{cr}}, &\qquad P<P_\mathrm{cr}
      \\
      0, &\qquad P>P_\mathrm{cr}     
    \end{array}
  \right.,
\end{align*}
where $P_\mathrm{cr} = \Delta E/(2\gamma)$.  It is simple to verify
that
\begin{align*}
  &I_+ \to 0,
  \\
  &I_- \to \frac{\pi \Delta E \mu^2}{\varepsilon_0 c \hbar k_B T} 
  \left\{
    \begin{array}{ll}
      P, &\qquad P<P_\mathrm{cr}
      \\
      P_\mathrm{cr}, &\qquad P>P_\mathrm{cr}
    \end{array}
  \right.,
\end{align*}
which is again the result for a single species characterized by the 
parameters $\Delta E$, $\mu$ and $\gamma$.

Due to line broadening, which in the microwave region of interest for
a $\mathrm{NH}_3$--$\mathrm{ND}_3$ mixture is mostly due to molecular
scattering, the absorption coefficient discussed above can not be
directly compared with that measured in an experiment.  However, a
test of the excitation energies (\ref{omegapm}) could be done in an
indirect way as in
\cite{Bleaney-Loubster.1948,Bleaney-Loubster.1950,Birnbaum-Maryott.1953b}.
In these experiments the spectrum of the absorption coefficient
measured for $\mathrm{NH}_3$ or $\mathrm{ND}_3$ at different pressures
was fitted by a Van Vleck--Weisskopf formula \cite{VVW} in which the
center line frequency is considered a fitting parameter. The results
of the fit are in excellent agreement with the predictions of our
model for a single chiral gas, see \cite{jptprl} or Appendix
\ref{inversion.line}.  In the case of a
$\mathrm{NH}_3$--$\mathrm{ND}_3$ mixture the situation is not
different. If our model is correct, with a 10\% $\mathrm{ND}_3$
fraction one should measure an absorption spectrum, mainly due to
the $\hbar\overline{\omega}_+$ excitation mode, see
Fig.~\ref{intensitieschimix1}, centered at a frequency which at high
pressures approaches the value of $0.26~\textrm{cm}^{-1}$, see
Fig.~\ref{inversionlineschimix1}. This is a striking difference with
respect to the case of a pure $\mathrm{NH}_3$ gas in which the center
line frequency of the absorption spectrum vanishes for $P>
1.69~\textrm{atm}$.  Also the behavior of
$\hbar\overline{\omega}_-(P)$ is rather different from the inversion
frequency of pure $\mathrm{ND}_3$. In fact, the two frequencies vanish
at $0.70~\textrm{atm}$ and $0.11~\textrm{atm}$, respectively.
However, this effect should be more difficult to measure as the
intensity $I_-$ is much lower than $I_+$, see
Fig.~\ref{intensitieschimix1}.

\section{Final remarks}

Our approach to the existence of chiral molecules is based on ideas of
equilibrium statistical mechanics. One may be surprised by the
presence of a quantum phase transition at room temperatures. We
emphasize that the transition takes place only in the inversion
degrees of freedom. The dynamics of these degrees of freedom is
affected by temperature only through the values of the coupling
constants.  A derivation of our approach from first principles is an
open question.  As we have discussed, in the case of a pure gas of
ammonia or deuterated ammonia, our model compares very well with
experimental results.

The problem of existence of chiral molecules has been approached also
from a dynamical point of view \cite{TH,BB,CGG} in terms of the
quantum linear evolution with a Lindblad term.  The question asked by
these authors is how long a molecule in a chiral localized state
remains in this state considering the interaction with the
environment.  This question had already been posed by Hund \cite{Hund}
without, however, any quantitative estimate.  In Refs. \cite{TH,BB}, a
calculation \textit{ab initio} is discussed exploiting a mechanism
whose basic idea goes back to \cite{Simonius,Harris-Stodolsky}:
through repeated scatterings in a host gas, a molecule is blocked in a
localized state.  A detailed analysis of the decoherence of a two-state
tunneling molecule due to collisions with a host gas is given in
\cite{CGG} in terms of a succession of quantum states of the molecule
satisfying the conditions for a consistent family of histories.

Effective nonlinearities, closer to our interpretation, were
considered in a dynamical context by \cite{Silbey-Harris} and
qualitatively connected to the disappearance of the inversion line.
However, no quantitative estimate was attempted.  In \cite{gsjpa}
dissipative effects are included in a nonlinear time-dependent model
obtaining qualitative conclusions similar to \cite{jptprl}.  What is
missing is a theory of the parameters of the model.

An important test of the present theory is to verify the shift of the
critical pressure depending on the percentage of the two species in a
binary mixture, e.g., $\mathrm{NH}_3$--$\mathrm{ND}_3$.  As we have
shown an addition of 10\% of $\mathrm{ND}_3$ is sufficient to halve
the critical pressure of pure $\mathrm{NH}_3$ and to appreciably change the
dependence on pressure of the inversion frequencies of both species.

The interpretation of the experimental data
\cite{Bleaney-Loubster.1948,Bleaney-Loubster.1950,Birnbaum-Maryott.1953b}
is based on the Van Vleck--Weisskopf formula.  The inversion line
measured at sufficiently high pressures consists actually of a band of
lines close to each other arising from different angular momentum
states.  At low pressures these lines can be resolved
\cite{Bleaney-Penrose.1947}.  We may ask the question whether this
case can be discussed in terms of a mixture of different pre-chiral
gases corresponding to ammonia molecules in different angular momentum
states and study the crossover to higher pressures.

\begin{acknowledgments}
  The authors would like to thank P. Facchi for very useful discussions
  in the early stages of this work.
\end{acknowledgments}

\appendix

\section{Stationary states of a chiral gas}
\label{stationarystateschiralgas}
The nonlinear eigenvalue problem associated with Eq.~(\ref{acca}) and
the normalization condition, namely,
\begin{align}
  \label{nlee}
    h(\psi_\lambda)\psi_\lambda = \lambda \psi_\lambda,
    \qquad \scp{\psi_\lambda}{\psi_\lambda}=N,
\end{align}
where
\begin{align*}
    h(\psi)=-\frac{\Delta E}{2}\sigma^x- \frac{G}{N}\scp{\psi}{\sigma^z
      \psi} \sigma^z,
\end{align*}
has different solutions depending on the value of the ratio $G/ \Delta
E$.  To see it, let us write
\begin{align*}
  \psi_\lambda = a_\lambda \varphi_+ + b_\lambda \varphi_-.
\end{align*}
The coefficients $a_\lambda$ and $b_\lambda$ can be chosen real and
are normalized to the number of molecules in the gas,
$a_\lambda^2+b_\lambda^2=N$.  By inserting this expression into
Eq.~(\ref{nlee}), we find that $a_\lambda,b_\lambda$ are the solutions
of the equation
\begin{align}
  \Delta E a_\lambda b_\lambda = \frac{G}{N} \left(
    a_\lambda^2-b_\lambda^2 \right) 2a_\lambda b_\lambda.
  \label{ab}
\end{align}
Once a solution of this system is found, the corresponding eigenvalue
is given by
\begin{align*}
  \lambda N = - \frac{\Delta E}{2} \left( a_\lambda^2-b_\lambda^2
  \right) - \frac{4G}{N} a_\lambda^2 b_\lambda^2.
\end{align*}

For any value of $G$, Eq.~(\ref{ab}) always admits solutions such that
$a_\lambda b_\lambda=0$.  Up to an irrelevant sign, there are two
different solutions which satisfy this condition and the normalization
rule, namely, the delocalized eigenstates
\begin{align}
  &\psi_{\lambda_1}=\sqrt{N} \varphi_+, \qquad \lambda_1=-\Delta E/2,
  \label{s1} \\
  &\psi_{\lambda_2}=\sqrt{N} \varphi_-, \qquad \lambda_2=+\Delta E/2.
  \label{s2}
\end{align}
If $G>\Delta E/2$, there appear two further solutions such that
$a_\lambda b_\lambda \neq 0$, namely, the roots of
\begin{align}
  2 a_\lambda^2/N - 1 = \Delta E/(2G).
  \label{s3s4}
\end{align}
The corresponding eigenstates are
\begin{align}
  &\psi_{\lambda_3}= \sqrt{N}\left( \sqrt{\frac{1}{2} +\frac{\Delta
        E}{4G}}~\varphi_+ +\sqrt{\frac{1}{2}-\frac{\Delta
        E}{4G}}~\varphi_-\right),
  \nonumber \\ &\lambda_3=-G, \label{s3}\\
  &\psi_{\lambda_4}= \sqrt{N}\left( \sqrt{\frac{1}{2} +\frac{\Delta
        E}{4G}}~\varphi_+ -\sqrt{\frac{1}{2}-\frac{\Delta
        E}{4G}}~\varphi_-\right). \nonumber \\ &\lambda_4=-G, \label{s4}
\end{align}
We call these states chiral in the sense that
$\psi_{\lambda_3}=\sigma^x \psi_{\lambda_4}$ and
$\psi_{\lambda_4}=\sigma^x \psi_{\lambda_3}$.  In the limit $G\gg
\Delta E$, the eigenstates $\psi_{\lambda_3}$ and $\psi_{\lambda_4}$
approach the localized states $\sqrt{N}\varphi_L$ and
$\sqrt{N}\varphi_R$, respectively.

\begin{figure}
  \includegraphics[width=\columnwidth,clip]{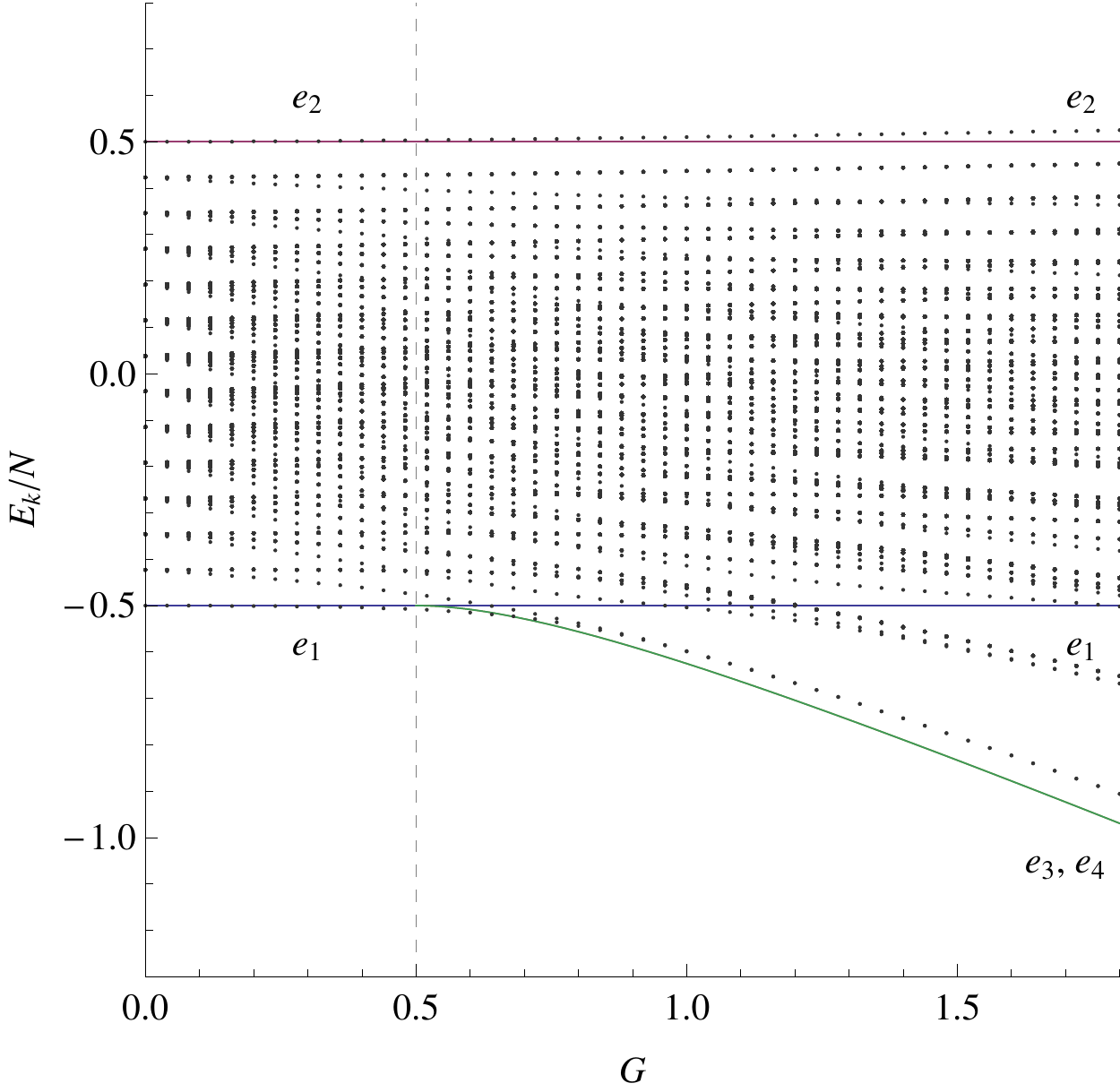}
  \centering
  \caption{(Color online) Eigenvalues $E_k$, $k=1,\dots,2^N$, divided
    by $N$, of the $N$-body Hamiltonian of Eq.~(\ref{HN}) with
    $\mathfrak{g}_{ij}=G/N$ and $N=13$ calculated numerically for
    different values of $G$ (dots).  The solid lines are the
    single-molecule energies $e_i=\mathcal{E}(\psi_{\lambda_i})/N$
    associated with the mean-field stationary states $\psi_{\lambda_i}$,
    $i=1,\dots,4$.}
  \label{spectrum}
\end{figure}

The states $\psi_\lambda$ determined above, are the stationary
solutions $\psi(t) = \exp(-i\lambda t/ \hbar) \psi_\lambda$ of the
time-dependent nonlinear Schr\"odinger equation
\begin{align*}
  i \hbar \frac{\d}{\d t} \psi(t) = h(\psi) \psi(t)
\end{align*}
and the eigenvalue $\lambda$ is the Lagrange multiplier fixing the
conservation of the number of molecules.  Associated with any solution
$\psi(t)$ of this equation there is a conserved energy given by
\begin{align*}
  \mathcal{E}(\psi) =
  -\frac{\Delta E}{2} \scp{\psi}{\sigma^x \psi}
  -\frac{G}{2N} \scp{\psi}{\sigma^z \psi}^2.
\end{align*}
The value of this functional calculated at the stationary solutions
(\ref{s1})--(\ref{s4}), i.e.,
\begin{align*}
  \mathcal{E}(\psi_{\lambda}) = -\frac{\Delta E}{2} \left(
    a_\lambda^2-b_\lambda^2 \right) - \frac{2G}{N} a_\lambda^2
  b_\lambda^2,
\end{align*}
provides the corresponding single-molecule energies
$e_i=\mathcal{E}(\psi_{\lambda_i})/N$, $i=1,\dots,4$,
\begin{align*}
  &e_1=-\frac{\Delta E}{2} 
  \nonumber\\
  &e_2= +\frac{\Delta E}{2} 
  \\ 
  &e_3=e_4= -\frac{\Delta E}{2}-\frac{1}{2G}
  \left( \frac{\Delta E}{2}-G \right)^2.
  \nonumber
\end{align*}
These energies are plotted in Fig.~\ref{spectrum} as a function of
$G$.  The molecular state effectively assumed by the gas is that with
the minimal energy, namely, the symmetric delocalized state
$\psi_{\lambda_1}$ for $G<\Delta E/2$, or one of the two degenerate
chiral states $\psi_{\lambda_3}$, $\psi_{\lambda_4}$ for $G>\Delta
E/2$.
                                   
The above results imply a bifurcation of the mean-field ground state
at a critical interaction strength $G_\mathrm{cr}=\Delta E/2$.
According to Eq.~(\ref{G}), this transition can be obtained for a
given molecular species by increasing the gas pressure above the
critical value $P_\mathrm{cr}$ given by Eq.~(\ref{pcr}).

\section{Excitations from the ground state}
\label{excitationsfromgroundstate}
It is interesting to compare the mean-field molecular states described
in the previous section with the eigenstates of the $N$-body model
(\ref{HN}).  In the latter case, the Hamiltonian is represented by a
$2^N \times 2^N$ matrix which, if $N$ is not too large, can be
diagonalized numerically.  For a meaningful comparison with the mean
field we set $\mathfrak{g}_{ij}=G/N$.  The eigenvalues obtained for
$N=13$ are shown in Fig.~\ref{spectrum} as a function of $G$.  We see
that there is a strict correspondence between the smallest and largest
eigenvalues of (\ref{HN}) and the smallest and largest conserved energies
associated with the mean-field eigenstates of (\ref{acca}).  Assuming
that the eigenstates of (\ref{HN}) are ordered according to their
magnitude, $E_1<E_2<\dots,E_{2^N}$, we have
\begin{align*}
  &E_1 \simeq \left\{
    \begin{array}{ll}
      \mathcal{E}(\psi_{\lambda_1}), &\quad G< \Delta E/2,      
      \\
      \mathcal{E}(\psi_{{\lambda_3},{\lambda_4}}), &\quad G>\Delta E/2,      
    \end{array}
  \right.
  \\
  &E_{2^N} \simeq \mathcal{E}(\psi_{\lambda_2}).
\end{align*}
By increasing the size $N$ of the system we find that the differences
between the $N$-body and mean-field energies indicated above increase
slower than $N$. Assuming that these differences are $o(N)$, in
Fig.~\ref{spectrum} we compare single-particle rescaled values, namely,
$E_k/N$, $k=1,\dots,2^N$, and $e_i=\mathcal{E}(\psi_{\lambda_i})/N$,
$i=1,\dots,4$.  The results shown for $N=13$ already give an idea of
what to expect for $N\to\infty$.

\begin{figure}
  \centering
  \includegraphics[width=\columnwidth,clip]{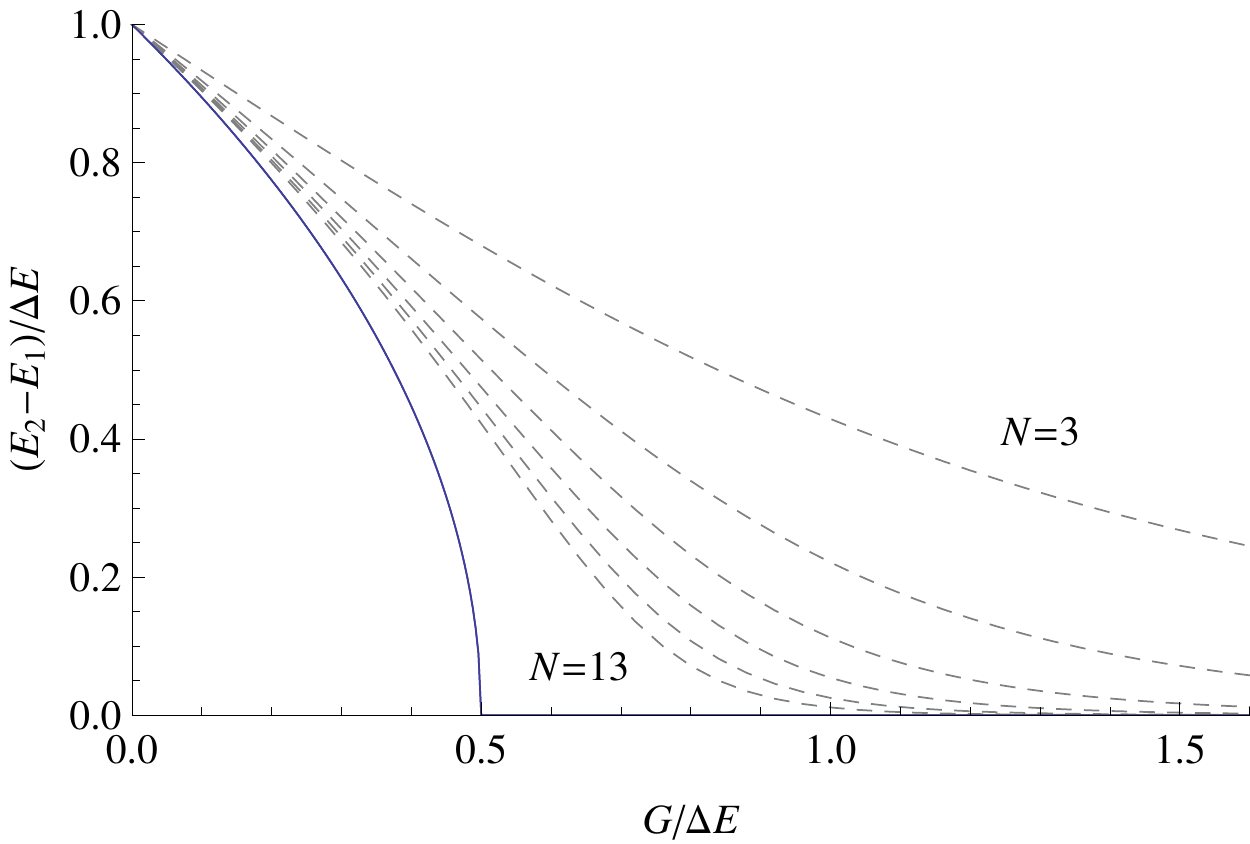}
  \caption{(Color online) Energy gap $E_2-E_1$ of the $N$-body
    Hamiltonian (\ref{HN}) with $\mathfrak{g}_{ij}=G/N$ and $N=3$, 5,
    7, 9, 11, 13 (dashed lines from top to bottom) as a function of
    the ratio $G/\Delta E$.  The solid line represents $\hbar
    \overline\nu/\Delta E$, where $\overline\nu$ is the inversion
    doubling frequency predicted by Eq.~(\ref{nubar}).}
  \label{gap}
\end{figure}

The diagonalization of the Hamiltonian (\ref{HN}) provides a full
description of the excitation modes. The energy gap between the ground
state and the first excited one, namely, $E_2-E_1$, is the inversion
doubling, namely, the excitation of one quasi-molecule from the state
$\varphi_+$ to the state $\varphi_-$. For $G=0$, the gap amounts
exactly to $\Delta E$ and the excited level is $N$-fold degenerate.
Due to the attractive dipole-dipole interaction, the value of the gap
evaluated for $G>0$ is decreased with respect to the noninteracting
case by an amount of the order of $G$.  For $G$ large the ground state
becomes two-fold degenerate and the gap vanishes.  In Fig.~\ref{gap}
we show the value of $E_2-E_1$ evaluated as a function of $G$ for
systems of increasing size $N$.
For any value of $N$ the gap vanishes smoothly at $G\gg \Delta E$,
however, by increasing $N$ such smoothness becomes paler and paler.
What is the behavior that we should expect in the $N\to\infty$ limit?
As we will see in a moment, the mean-field indicates that $E_2-E_1$
sharply vanishes at $G=\Delta E/2$ thus suggesting that for $N\to
\infty$ a phase transition takes place.

In general, the excitation energies from the ground state can be
evaluated as the poles of the Fourier transform of the linear-response
function \cite{Blaizot-Ripka}. 
In \cite{jptprl}, we have evaluated the linear response of the 
mean-field Hamiltonian (\ref{acca}) to a time-dependent perturbation 
which corresponds to an electro-magnetic radiation coupling with
the electric dipole of the molecules. 
For a gas in the mean-field $\psi_{\lambda_1}$, the Fourier transform
of the linear-response function is \cite{noteRCHI}
\begin{align}
  \mathcal{R}(\omega) = \frac{2\Delta E}{(\hbar \omega)^2-(\Delta
    E^2-2G\Delta E)},
  \label{linear.response.prl}
\end{align}
which has two simple poles at angular frequency
$\pm\overline{\omega}$, $\overline{\omega}=\sqrt{\Delta E^2-2G\Delta
  E}/\hbar$.  We conclude that, in the region $0\leq G <\Delta E/2$
where the gas is at equilibrium in the molecular state
$\psi_{\lambda_1}$, the mean-field model predicts an inversion
doubling frequency $\overline{\nu}=\overline{\omega}/(2\pi)$ given by
\begin{align}
  \overline{\nu}= \frac{\Delta E}{h} \sqrt{1-\frac{2G}{\Delta E}},
  \label{nubar}
\end{align}
which vanishes at $G=\Delta E/2$.  This behavior of $h \overline{\nu}$
is compatible with the $N\to\infty$ limit of the gap $E_2-E_1$ which
can be envisaged from Fig.~\ref{gap}.

\section{Pressure dependence of $\mathrm{NH}_3$ and $\mathrm{ND}_3$
  inversion lines: comparison with experiments}
\label{inversion.line}
\begin{figure}
  \centering
  \includegraphics[width=\columnwidth,clip]{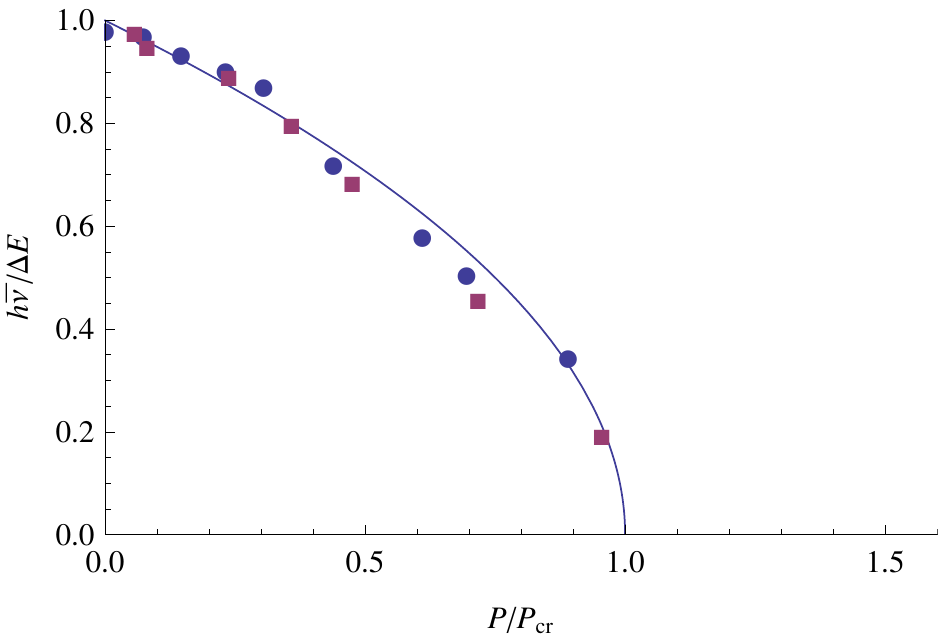}
  \caption{(Color online) Plot of the universal function of
    Eq.~(\ref{unifunapp}).  Points are the experimental results for
    $\mathrm{NH}_3$ (circles) or $\mathrm{ND}_3$ (squares) available
    from Ref. \cite{Bleaney-Loubster.1948,Bleaney-Loubster.1950} and
    \cite{Birnbaum-Maryott.1953b}, respectively.  The experimental
    data have been scaled using the values of $\Delta E$ and
    $P_\mathrm{cr}$ from Table \ref{pcr_tab}.}.
  \label{universal_line}
\end{figure}
In \cite{jptprl}, we have compared the mean-field theoretical
prediction for the inversion frequency with the spectroscopic
data available for ammonia
\cite{Bleaney-Loubster.1948,Bleaney-Loubster.1950} and deuterated
ammonia \cite{Birnbaum-Maryott.1953b}.  In these experiments, the
central frequency of the inversion peak was measured at different
pressures. The frequency is found to decrease by
increasing $P$ and vanishes for pressures greater than a critical
value.  This behavior is very well accounted for by the mean-field
prediction (\ref{nubar}) for $\overline{\nu}$.  In fact, an important
feature of this equation is the statement that the ratio between the
frequency $\overline{\nu}$ of a measured spectroscopic inversion line
and the energy splitting $\Delta E$ of an isolated molecule,
\begin{align}
  \frac{h \overline{\nu}}{\Delta E} =
  \sqrt{1-\frac{P}{P_\mathrm{cr}}},
  \label{unifunapp}
\end{align}
is a universal function of $P/P_\mathrm{cr}$ independent of the
kind of chiral species considered. In Fig.~\ref{universal_line}, we
show the values of $h \overline{\nu}/\Delta E$ measured at different
pressures $P$ in the case of ammonia and deuterated ammonia.  The
superposition of the two sets of data plotted as a function of
$P/P_\mathrm{cr}$ and the agreement with formula (\ref{unifunapp}) is
impressive.

Some comments are in order.  The mean-field prediction
(\ref{unifunapp}) does not contain free parameters.  The value of
$P_\mathrm{cr}$ is evaluated from the electric dipole $\mu$ of the
molecule, its collision diameter $d$ and inversion energy splitting
$\Delta E$, and the temperature $T$ of the gas, see Eq.~(\ref{pcr})
and Table \ref{pcr_tab}.  We remark, however, that the experimental
points shown in Fig.~\ref{universal_line} are the result of an
indirect measurement.  More precisely, in experiments
\cite{Bleaney-Loubster.1948,Bleaney-Loubster.1950,Birnbaum-Maryott.1953b}
the line shape of the absorption coefficient measured for
$\mathrm{NH}_3$ and $\mathrm{ND}_3$ at different pressures $P$ is
fitted by a Van Vleck--Weisskopf formula \cite{VVW} in which
$\overline{\nu}$ is a considered a fitting parameter.


\begin{thebibliography}{99}
\bibitem{note.units} We use SI units. The expression of $\chi(\omega)$
  in the Gauss system of units is obtained by changing $4\pi
  \varepsilon_0 \to 1$.
\bibitem{nota.gaspuro}For a single chiral gas in the localized phase we
  have $2a^2/N-1=\Delta E/(2G)$, see Eq.~(\ref{s3s4}).
\bibitem{noteRCHI} The linear-response function $\mathcal{R}(\omega)$
  coincides up to a constant, namely, $\rho\mu^2/\varepsilon_0$, with
  the single-species limit of the susceptibility $\chi(\omega)$
  determined in Section \ref{inversionfrequencies.chimix}. Note that
  $\mathcal{R}(\omega)$ has been evaluated in \cite{jptprl} only in
  the delocalized phase $G < \Delta E/2$, while $\chi(\omega)$ is
  given also in the localized phase.
\end{thebibliography}
\end{document}